\documentclass[11pt,a4paper]{article}
\pdfoutput=1
\usepackage{jheppub}

\usepackage{url}
\usepackage{bm}

\allowdisplaybreaks[2]

\def\cD{\mathcal{D}}

\def\cN{\mathcal{N}}

\def\mint{\int_{-\infty}^\infty\!\cdots\!\int_{-\infty}^\infty}

\newcommand{\be}{\begin{equation}}
\newcommand{\ee}{\end{equation}}
\newcommand{\ba}{\begin{aligned}}
\newcommand{\ea}{\end{aligned}}

\def\vev#1{\langle #1 \rangle}

\def\({\left(}
\def\){\right)}

\DeclareMathOperator{\Tr}{Tr}

\preprint{RUP-25-6}

\title{Deformed Schur indices and Macdonald polynomials}

\author{Yasuyuki Hatsuda}

\affiliation{Department of Physics, Rikkyo University, Toshima, Tokyo 171-8501, Japan}

\emailAdd{yhatsuda@rikkyo.ac.jp}

\abstract{
The Schur index in four-dimensional $\mathcal{N}=4$ super Yang-Mills theory with $U(N)$ gauge group has a natural two-parameter deformation.
We find that a matrix integral in such a deformed Schur index can be exactly evaluated by using Macdonald polynomials. The resulting expression is a simple combinatorial summation over partitions. 
An extension to line operator indices is straightforward. In particular, for an anti-symmetric representation, the line operator index has a relatively simple form. We further discuss infinite $N$ analysis and finite $N$ giant graviton expansions. 
}

\dedicated{Dedicated to the memory of Masatoshi Noumi}

\begin{document}

\maketitle

\renewcommand{\thefootnote}{\arabic{footnote}}
\setcounter{footnote}{0}
\setcounter{section}{0}

\section{Introduction}
Superconformal indices \cite{Romelsberger:2005eg, Kinney:2005ej} are a powerful tool to probe BPS spectra in superconformal field theories.
They can be used to test the AdS/CFT correspondence as well as to study strongly coupled non-Lagrangian theories. In four-dimensional $\mathcal{N}=4$ super Yang-Mills theory with $U(N)$ gauge group, the superconformal index is well-known, and results in a unitary matrix integral \cite{Kinney:2005ej}. 

The $\mathcal{N}=4$ superconformal index has four independent fugacities. There are many specializations.
In this work, we focus on one of them, in which three of four independently remain.
More concretely, we analyze the following reduced index:
\begin{equation}
\begin{aligned}
I_N(t, u;q)=\int_{U(N)} dU  \exp \left( \sum_{n=1}^\infty \frac{1}{n}\frac{t^n+u^n-t^n u^n-q^n}{1-q^n} \Tr (U^n) \Tr [ (U^\dagger)^n ] \right),
\label{eq:Schur-0-1}
\end{aligned}
\end{equation}
where $dU$ is the (normalized) Haar measure of $U(N)$. 
An advantage of this reduction is that the integrand can be rewritten in terms of the $q$-Pochhammer symbol, while the original superconformal index is rewritten in terms of the elliptic gamma function. The analysis of the former gets simpler.
On the one hand this index is regarded as a reduction of the full superconformal index, but on the other hand it is also regarded as a two-parameter deformation of the Schur index \cite{Gadde:2011ik, Gadde:2011uv}, which is a very special limit of the superconformal index.
We refer to $I_N(t, u;q)$ as the \textit{deformed Schur index}.\footnote{From the perspective in the next section, it seems natural to refer to it as ``Macdonald index''. However this terminology has already been used in a similar but slightly different context in \cite{Gadde:2011uv}. To avoid confusion, we do not use this terminology.}
Note that the similar limit was considered (but not analyzed in detail) for other gauge groups in \cite{Spiridonov:2010qv}.

Clearly, the deformed Schur index \eqref{eq:Schur-0-1} also takes the form of the unitary matrix integral. Its evaluation is still far from obvious.
A powerful approach to evaluate it is the character expansion method \cite{Dolan:2007rq}.
Another approach is the so-called Fermi-gas formalism \cite{Gaiotto:2020vqj, Gaiotto:2021xce, Hatsuda:2022xdv}.\footnote{Precisely speaking, to apply the Fermi-gas formalism, we need an additional constraint that $q=tu$.}
In this work, we find the third approach based on less familiar symmetric orthogonal polynomials, \textit{Macdonald polynomials}.
By using them, the unitary matrix integral \eqref{eq:Schur-0-1} can be directly performed for arbitrary $N$. We find the following surprisingly simple sum representation:
\begin{equation}
\begin{aligned}
I_N(t, u; q)=\frac{(q;q)_\infty}{(t;t)_{N}(t^Nq;q)_\infty}\sum_{\ell(\lambda) \leq N} u^{|\lambda|}\prod_{i=1}^{\ell(\lambda)} \frac{(t^{N-i+1};q)_{\lambda_i}}{(t^{N-i}q;q)_{\lambda_i}}.
\end{aligned}
\label{eq:Schur-0-2}
\end{equation}
where $\lambda$ is a partition whose length $\ell(\lambda)$ is less than or equal to $N$ (see Appendix~\ref{app:Mac}).
We stress that the result \eqref{eq:Schur-0-2} is perturbative in $u$ but exact in $t$ and $q$.
As a consequence, it is particularly useful in the study of finite $N$ corrections, also known as giant graviton expansions \cite{Arai:2019xmp, Arai:2020qaj, Imamura:2021ytr, Gaiotto:2021xce}.
A similar evaluation is possible for line operator indices, in which we insert characters of the $U(N)$ gauge group into \eqref{eq:Schur-0-1}.

The organization of this paper is as follow. In the next section, we derive \eqref{eq:Schur-0-2} by using the Macdonald polynomials. We use known mathematical results on the Macdonald polynomials. All the ingredients needed in this work are reviewed in Appendix~\ref{app:Mac}. We consider various special limits, and confirm the validity of \eqref{eq:Schur-0-2}. We also comment on advantages of our result \eqref{eq:Schur-0-2}, compared to the result obtained by the character expansion method. We can extend the similar computation to line operator insertions. In particular, the line operator index for an anti-symmetric representation  has a simple expression. In Section~\ref{sec:large-N}, we propose a new systematic way to deal with the large $N$ analysis. We also explore finite $N$ corrections to the indices. We find some new analytic results on the giant graviton expansions. Section~\ref{sec:conclusion} is devoted to future directions. In Appendix~\ref{sec:interface}, we show additional results on half-indices of interfaces with $U(N)$ and $U(M)$ gauge groups.

\section{Deformed Schur indices from Macdonald polynomials}\label{sec:Schur}
\subsection{Superconformal indices and various limits}
We start with a matrix integral representation of the $\cN=4$ superconformal index for the $U(N)$ gauge group. As shown in \cite{Kinney:2005ej}, it is given by
\begin{equation}
\begin{aligned}
I_N(t, u, v;p, q)=\int_{U(N)} dU  \exp \left( \sum_{n=1}^\infty \frac{f(t^n, u^n, v^n; p^n, q^n)}{n} \Tr (U^n) \Tr [ (U^\dagger)^n ] \right),
\label{eq:SCI}
\end{aligned}
\end{equation}
where $f(t, u, v;p, q)$ is the single letter index of the theory. In our convention, it is given by
\begin{equation}
\begin{aligned}
f(t, u, v; p, q)=1-\frac{(1-t)(1-u)(1-v)}{(1-p)(1-q)}.
\end{aligned}
\label{eq:single-letter-index}
\end{equation}
It is important to note that five parameters $(t, u, v; p, q)$ are not independent. We have a constraint
\begin{equation}
\begin{aligned}
pq=tuv. 
\label{eq:constraint}
\end{aligned}
\end{equation}
Therefore four of five are actually independent.
 Since the integrand in \eqref{eq:SCI} is a class function of the unitary matrix $U$, we can use Weyl's integration formula (see \cite{Fulton} for instance). It reduces to an integral over the maximal torus $\mathbb{T}^N$. The resulting integral takes the form
\begin{equation}
\begin{aligned}
I_N(t, u, v; p, q)&=\frac{1}{N!} \oint_{\mathbb{T}^N} \prod_{i=1}^N \frac{dx_i}{2\pi i x_i} \prod_{1 \leq i \ne j \leq N} \left( 1-\frac{x_i}{x_j} \right)\\
&\quad \times \exp \left( \sum_{n=1}^\infty \frac{f(t^n, u^n, v^n; p^n, q^n)}{n} p_n(x) p_n(x^{-1}) \right),
\end{aligned}
\label{eq:SCI-2}
\end{equation}
where $x=(x_1,\dots, x_N)$ are eigenvalues of $U$, and $p_n(x)$ is the power sum symmetric polynomial (see Appendix~\ref{app:Mac}). The integration contour for each $x_i$ goes around a unit circle counterclockwise.

It is not easy to perform this $N$-dimensional integral for the full superconformal index exactly. Instead, we explore a special case in the fugacity configuration so that the exact evaluation is possible. In this work, we focus on the following slice of the fugacity space:
\begin{equation}
\begin{aligned}
v=p=0.
\end{aligned}
\end{equation}
In this special case, the fugacity constraint \eqref{eq:constraint} is automatically satisfied, and we can independently change $(t, u, q)$.
The resulting single letter index is now reduced to
\begin{equation}
\begin{aligned}
f(t, u; q)=1-\frac{(1-t)(1-u)}{1-q}=\frac{t+u-tu-q}{1-q}.
\end{aligned}
\end{equation}
The matrix integral is thus given by \eqref{eq:Schur-0-1}.
There are several interesting specializations of this index. 

For $t=q$ or $u=q$, things are dramatically simplified. In this case, the single letter index reduces to $f(q, u;q)=u$ or $f(t, q;q)=t$, and it does not depend on $q$.
In this very special limit, the index counts the $1/2$ BPS operators. It is well-known that the index is exactly given by
\begin{equation}
\begin{aligned}
I_N(q, u; q)=\frac{1}{(u;u)_N},\qquad I_N(t, q;q)=\frac{1}{(t;t)_N}.
\end{aligned}
\end{equation}

If taking the limit $q\to 0$, we have $f(t,u;0)=t+u-tu$. This case counts $1/4$ BPS operators. 
The index in this limit is more non-trivial than the $1/2$ BPS case.
We will derive an exact form of its index in the next subsection.

Also taking the limit $u\to 0$, the single letter index is given by
\begin{equation}
\begin{aligned}
f(t,0;q)=\frac{t-q}{1-q}.
\end{aligned}
\end{equation}
In this case, the corresponding index is known as the ``half-index'' (of Neumann boundary condition) \cite{Gaiotto:2019jvo}. 
Due to an obvious symmetry between $t$ and $u$, the limit $t \to 0$ is essentially same as $u \to 0$.
However, in the analysis in the next subsection, these two limits look different, and lead to the equivalent result non-trivially.

Finally, if we set $u=q/t$, the resulting index is known as the flavored Schur index. The single letter index is now given by
\begin{equation}
\begin{aligned}
f(t,q/t;q)=\frac{t+q/t-2q}{1-q}.
\end{aligned}
\end{equation}
For the further specialization to $t=q^{1/2}$ (\textit{i.e.}, $u=q^{1/2}$), the index $I_N(q^{1/2}, q^{1/2}; q)$ is nothing but the original Schur index \cite{Gadde:2011ik, Gadde:2011uv}. 

We stress that all of them are obtained from the index \eqref{eq:deformed-Schur} as special limits.
The reduced index $I_N(t, u; q)$ is regarded as a two-parameter deformation of the Schur index $I_N(q^{1/2}, q^{1/2}; q)$. As mentioned in the introductory section, we refer to $I_N(t, u; q)$ as the deformed Schur index.

\subsection{Exact evaluation of deformed Schur indices}
In this subsection, we evaluate the matrix integral of the deformed Schur index exactly.
When $v=p=0$, we can rewrite the integral representation \eqref{eq:SCI-2} as a more convenient form in terms of the $q$-Pochhammer symbol:
\begin{equation}
\begin{aligned}
I_N(t, u; q)&=\frac{1}{N!}\frac{(q;q)_\infty^{N}(tu;q)_\infty^N}{(t;q)_\infty^N (u;q)_\infty^N} \oint_{\mathbb{T}^N} \prod_{i=1}^N \frac{dx_i}{2\pi i x_i} \prod_{1\leq i \ne j \leq N} \frac{(x_i/x_j;q)_\infty (t u x_i/x_j;q)_\infty}{(tx_i/x_j;q)_\infty (ux_i/x_j;q)_\infty}.
\end{aligned}
\label{eq:deformed-Schur}
\end{equation}
This is a starting point of our analysis. The $q$-Pochhammer symbol is defined by
\begin{equation}
\begin{aligned}
(x;q)_\infty=\prod_{k=0}^\infty (1-xq^k),\qquad (x;q)_n=\prod_{k=0}^{n-1}(1-xq^k),\qquad
(x;q)_0=1,
\end{aligned}
\end{equation}
and we have used an identity,
\begin{equation}
\begin{aligned}
(qx;q)_\infty=\frac{(x;q)_\infty}{1-x},
\end{aligned}
\label{eq:q-Poch-id-1}
\end{equation}
to derive \eqref{eq:deformed-Schur}.
A method to perform the integral \eqref{eq:deformed-Schur} is simple. The computation consists of three steps.

In the first step, we recognize that the integrand of \eqref{eq:deformed-Schur} includes a weight function of Macdonald polynomials of type $A_{N-1}$. In Appendix~\ref{app:Mac}, we review basics on the Macdonald polynomials of type A, based on \cite{Macdonald, Noumi},  for the reader's convenience. The weight function of the Macdonald polynomials of type $A_{N-1}$ is given by
\begin{equation}
\begin{aligned}
w(x)=\prod_{1\leq i \ne j \leq N}\frac{(x_i/x_j;q)_\infty}{(tx_i/x_j;q)_\infty}.
\end{aligned}
\end{equation}
The integral \eqref{eq:deformed-Schur} is then written as
\begin{equation}
\begin{aligned}
I_N(t, u; q)=\frac{1}{N!}\frac{(q;q)_\infty^{N}}{(t;q)_\infty^N} \oint_{\mathbb{T}^N} \prod_{i=1}^N \frac{dx_i}{2\pi i x_i} w(x) \prod_{i,j=1}^N \frac{(t u x_i/x_j;q)_\infty}{(ux_i/x_j;q)_\infty}.
\end{aligned}
\end{equation}

In the second step, we use the Cauchy formula for the Macdonald polynomials:
\begin{align}
\prod_{i,j=1}^N \frac{(t x_i y_j;q)_\infty}{(x_i y_j;q)_\infty}=\sum_{\ell(\lambda) \leq N} b_\lambda P_\lambda(x;q,t)P_\lambda(y;q,t),
\label{eq:Cauchy}
\end{align}
where
\begin{align}
b_\lambda=\prod_{1 \leq i \leq j \leq \ell(\lambda)} \frac{(t^{j-i+1}q^{\lambda_i-\lambda_j};q)_{\lambda_j-\lambda_{j+1}}}{(t^{j-i}q^{\lambda_i-\lambda_j+1};q)_{\lambda_j-\lambda_{j+1}}}.
\end{align}
The summation in \eqref{eq:Cauchy} is taken over all the partitions whose length is less than or equal to $N$.
Setting $y_j=u/x_j$, we obtain
\begin{equation}
\begin{aligned}
\prod_{i,j=1}^N \frac{(tu x_i/x_j;q)_\infty}{(ux_i/x_j;q)_\infty}=\sum_{\ell(\lambda) \leq N}u^{|\lambda|} b_\lambda P_\lambda(x;q,t)P_\lambda(x^{-1};q,t),
\end{aligned}
\end{equation}
where we have used $P_\lambda(ux^{-1};q,t)=u^{|\lambda|}P_\lambda(x^{-1};q,t)$.

In the final step, we perform the integral by using the norm formula of the Macdonald polynomial:
\begin{equation}
\begin{aligned}
\frac{1}{N!}\oint_{\mathbb{T}^N} \prod_{i=1}^N \frac{dx_i}{2\pi i x_i} w(x) P_\lambda(x;q,t)P_\lambda(x^{-1};q,t)=\cN_{\lambda,N},
\end{aligned}
\end{equation}
where
\begin{equation}
\begin{aligned}
\cN_{\lambda,N}=\prod_{1 \leq i<j \leq N} \frac{(t^{j-i}q^{\lambda_i-\lambda_j+1};q)_\infty (t^{j-i}q^{\lambda_i-\lambda_j};q)_\infty}{(t^{j-i+1}q^{\lambda_i-\lambda_j};q)_\infty (t^{j-i-1}q^{\lambda_i-\lambda_j+1};q)_\infty}.
\end{aligned}
\end{equation}
We arrive at an exact expression
\begin{equation}
\begin{aligned}
I_N(t, u; q)=\frac{(q;q)_\infty^{N}}{(t;q)_\infty^N}\sum_{\ell(\lambda) \leq N}u^{|\lambda|} b_\lambda \cN_{\lambda,N}.
\end{aligned}
\end{equation}
This result is already simple compared with the original integral expression. After some computations, we can further simplify the product $b_\lambda \cN_{\lambda,N}$ as
\begin{equation}
\begin{aligned}
b_\lambda \cN_{\lambda,N}=\frac{(t;q)_\infty^N}{(q;q)_\infty^{N-1}(t;t)_{N}(t^Nq;q)_\infty}\prod_{i=1}^{\ell(\lambda)} \frac{(t^{N-i+1};q)_{\lambda_i}}{(t^{N-i}q;q)_{\lambda_i}}.
\end{aligned}
\label{eq:bN}
\end{equation}
Therefore we find a more compact form:
\begin{equation}
\begin{aligned}
I_N(t, u; q)=\frac{(q;q)_\infty}{(t;t)_{N}(t^Nq;q)_\infty}\sum_{\ell(\lambda) \leq N} u^{|\lambda|}\prod_{i=1}^{\ell(\lambda)} \frac{(t^{N-i+1};q)_{\lambda_i}}{(t^{N-i}q;q)_{\lambda_i}}.
\end{aligned}
\label{eq:main}
\end{equation}
This is one of the main results in this work. 
In the original integral \eqref{eq:deformed-Schur}, the symmetric structure for $t$ and $u$ is manifest. However, our result \eqref{eq:main} is not. The symmetry is cured in a quite non-trivial way. This fact causes some interesting consequences.

To check the validity of our result \eqref{eq:main}, let us take various limits, mentioned in the previous subsection.
We first consider $t=q$. 
In this limit, the Macdonald polynomials reduce to the Schur polynomials. In fact, the integral \eqref{eq:deformed-Schur} is written as
\begin{equation}
\begin{aligned}
I_N(q, u; q)=\frac{1}{N!}\oint_{\mathbb{T}^N} \prod_{i=1}^N \frac{dx_i}{2\pi i x_i} \prod_{1\leq i \ne j \leq N} \left(1-\frac{x_i}{x_j} \right) \prod_{i,j=1}^N \frac{1}{1-ux_i/x_j},
\end{aligned}
\end{equation}
and we can use the more familiar Cauchy formula for the Schur polynomials:
\begin{equation}
\begin{aligned}
\prod_{i,j=1}^N \frac{1}{1-x_i y_j}=\sum_{\ell(\lambda) \leq N}s_\lambda(x)s_\lambda(y).
\end{aligned}
\end{equation}
The prefactor and the summand in \eqref{eq:main} get trivial, and as was shown in \cite{Dolan:2007rq}, the index finally becomes
\begin{equation}
\begin{aligned}
I_N(q, u; q)=\sum_{\ell(\lambda) \leq N} u^{|\lambda|}=\frac{1}{(u;u)_N},
\end{aligned}
\end{equation}
as expected. As we mentioned before, the deformed Schur index has the symmetry between $t$ and $u$. This means that for $u=q$, we have
\begin{equation}
\begin{aligned}
I_N(t,q;q)=\frac{(q;q)_\infty}{(t;t)_{N}(t^Nq;q)_\infty}\sum_{\ell(\lambda) \leq N} q^{|\lambda|}\prod_{i=1}^{\ell(\lambda)} \frac{(t^{N-i+1};q)_{\lambda_i}}{(t^{N-i}q;q)_{\lambda_i}}=\frac{1}{(t;t)_N}.
\end{aligned}
\end{equation}
Therefore we obtain a quite non-trivial summation identity
\begin{equation}
\begin{aligned}
\sum_{\ell(\lambda) \leq N} q^{|\lambda|}\prod_{i=1}^{\ell(\lambda)} \frac{(t^{N-i+1};q)_{\lambda_i}}{(t^{N-i}q;q)_{\lambda_i}}
=\frac{(t^Nq;q)_\infty}{(q;q)_\infty}.
\end{aligned}
\end{equation}
We do not have a direct proof of this identity. It would be nice to find it.

Let us consider the limit $q\to 0$. In this case, the Macdonald polynomials reduce to the Hall-Littlewood polynomials.
From \eqref{eq:main}, we have
\begin{equation}
\begin{aligned}
I_N(t, u; 0)&=\frac{1}{(t;t)_{N}}\sum_{\ell(\lambda) \leq N} u^{|\lambda|}\prod_{i=1}^{\ell(\lambda)} (1-t^{N-i+1}) \\
&=\frac{1}{(t;t)_N}\sum_{\ell(\lambda) \leq N} u^{|\lambda|}(t^N;t^{-1})_{\ell(\lambda)}.
\end{aligned}
\end{equation}
We can perform the sum as follows.
\begin{equation}
\begin{aligned}
\sum_{\ell(\lambda) \leq N} u^{|\lambda|}(t^N;t^{-1})_{\ell(\lambda)} =1+\sum_{\ell=1}^N (t^N;t^{-1})_\ell \sum_{\lambda_1 \geq \cdots \geq \lambda_\ell \geq 1} u^{\lambda_1+\cdots+\lambda_\ell}
\end{aligned}
\end{equation}
By changing the variables as
\begin{equation}
\begin{aligned}
\lambda_1=n_1+n_2+\cdots+n_\ell,\qquad
\lambda_2=n_2+\cdots+n_\ell,\qquad \dots,\qquad
\lambda_\ell=n_\ell,
\end{aligned}
\label{eq:lambda-n}
\end{equation}
we find
\begin{equation}
\begin{aligned}
\sum_{\lambda_1 \geq \cdots \geq \lambda_\ell \geq 1} u^{\lambda_1+\cdots+\lambda_\ell}
&=\sum_{n_1=0}^\infty\sum_{n_2=0}^\infty \cdots \sum_{n_{\ell-1}=0}^\infty \sum_{n_\ell=1}^\infty u^{n_1+2n_2+\cdots +(\ell-1)n_{\ell-1}+\ell n_\ell}\\
&=\sum_{n_1=0}^\infty u^{n_1} \sum_{n_2=0}^\infty u^{2n_2}\cdots \sum_{n_{\ell-1}=0}^\infty u^{(\ell-1)n_{\ell-1}} \sum_{n_\ell=1}^\infty u^{\ell n_\ell} \\
&=\frac{u^\ell}{(1-u)(1-u^2)\cdots (1-u^\ell)}\\
&=\frac{u^\ell}{(u;u)_\ell}
\end{aligned}
\end{equation}
We also have
\begin{equation}
\begin{aligned}
(t^N;t^{-1})_\ell=\frac{(t;t)_N}{(t;t)_{N-\ell}}.
\end{aligned}
\end{equation}
Combining these results, we finally arrive at a very simple expression of the $1/4$ BPS index,
\begin{equation}
\begin{aligned}
I_N(t, u; 0)&=\sum_{\ell=0}^N \frac{u^\ell}{(t;t)_{N-\ell} (u;u)_\ell}.
\end{aligned}
\label{eq:quarter}
\end{equation}
The symmetric structure is not manifest, but one can confirm it.

The limit $u \to 0$ is also interesting. The integral \eqref{eq:deformed-Schur} reduces to
\begin{equation}
\begin{aligned}
I_N(t, 0; q)&=\frac{1}{N!}\frac{(q;q)_\infty^{N}}{(t;q)_\infty^N} \oint_{\mathbb{T}^N} \prod_{i=1}^N \frac{dx_i}{2\pi i x_i} \prod_{1\leq i \ne j \leq N} \frac{(x_i/x_j;q)_\infty}{(tx_i/x_j;q)_\infty}.
\end{aligned}
\end{equation}
We can directly apply the norm formula for the trivial Macdonald polynomial $P_\emptyset(x;q,t)=1$. 
\begin{equation}
\begin{aligned}
I_N(t, 0; q)&=\frac{(q;q)_\infty^{N}}{(t;q)_\infty^N}\cN_{\emptyset,N}=\frac{(q;q)_\infty}{(t;t)_{N} (t^Nq;q)_\infty}.
\end{aligned}
\label{eq:half-I}
\end{equation}
where we have used \eqref{eq:bN} for $\lambda=\emptyset$ and $b_{\emptyset}=1$.
This is an exact result on the half-index.%
\footnote{Note that an essentially equivalent analytic result was obtained in \cite{Spiridonov:2010qv}. The same result is also observed in~\cite{Gaiotto:2019jvo} by using a duality on half-indices of Neumann boundary condition and Nahm pole boundary condition.}
Of course, the result agrees with \eqref{eq:main} for $u=0$. 
On the other hand, if we consider $t \to 0$, the Macdonald polynomials reduce to the $q$-Whittaker polynomials.
The sum in \eqref{eq:main} remains non-trivial,
\begin{equation}
\begin{aligned}
I_N(0, u;q)=(q;q)_\infty \sum_{\ell(\lambda) \leq N}  \frac{u^{|\lambda|}}{(q;q)_{\lambda_N}}.
\end{aligned}
\end{equation}
The symmetry between $t$ and $u$ requires $I_N(0, u;q)=I_N(u, 0;q)$, and
we get a sum formula,
\begin{equation}
\begin{aligned}
\sum_{\ell(\lambda) \leq N}  \frac{u^{|\lambda|}}{(q;q)_{\lambda_N}}=\frac{1}{(u;u)_{N} (u^Nq;q)_\infty}.
\end{aligned}
\end{equation}
 We can show it directly as follows. Changing the variables as in \eqref{eq:lambda-n} for $\ell=N$, we have
 \begin{equation}
\begin{aligned}
\sum_{\ell(\lambda) \leq N}  \frac{u^{|\lambda|}}{(q;q)_{\lambda_N}}
&=\sum_{n_1=0}^\infty\cdots \sum_{n_N=0}^\infty \frac{u^{n_1+2n_2+\cdots+N n_N}}{(q;q)_{n_N}}\\
&=\frac{1}{(1-u)(1-u^2)\cdots (1-u^{N-1})}\frac{1}{(u^N;q)_\infty}\\
&=\frac{1}{(u;u)_{N-1} (u^N;q)_\infty}=\frac{1}{(u;u)_{N} (u^Nq;q)_\infty}.
\end{aligned}
\end{equation}
All of these computations validate the result \eqref{eq:main}.

Let us consider the Schur limit $(t,u,q) \to (x,x,x^2)$. In this case, \eqref{eq:main} reduces to
\begin{equation}
\begin{aligned}
I_N(x,x;x^2)=\frac{(x^2;x^2)_\infty}{(x;x)_{N} (x^{N+2};x^2)_\infty}
\sum_{\ell(\lambda)\leq N} x^{|\lambda|} \prod_{i=1}^{\ell(\lambda)} \frac{(x^{N-i+1};x^2)_{\lambda_i}}{(x^{N-i+2};x^2)_{\lambda_i}}.
\end{aligned}
\end{equation}
Unfortunately, we do not have a way to deal with this summation over $\lambda$. Interestingly, there is another much simpler formula for the Schur index \cite{Bourdier:2015wda},
\begin{equation}
\begin{aligned}
I_N(x,x;x^2)=\frac{(x^2;x^2)_\infty}{(x;x)_\infty^2}\sum_{n=0}^\infty (-1)^n \biggl[ \binom{N+n}{N}+\binom{N+n-1}{N} \biggr] x^{nN+n^2}.
\end{aligned}
\end{equation}
Combining these two expressions, the following identity should hold,
\begin{equation}
\begin{aligned}
&\frac{(x;x)_\infty (x^{N};x)_\infty}{(x^{N};x^2)_\infty}\sum_{\ell(\lambda)\leq N} x^{|\lambda|} \prod_{i=1}^{\ell(\lambda)} \frac{(x^{N-i+1};x^2)_{\lambda_i}}{(x^{N-i+2};x^2)_{\lambda_i}}\\
&\qquad\qquad\qquad=\sum_{n=0}^\infty (-1)^n \biggl[ \binom{N+n}{N}+\binom{N+n-1}{N} \biggr] x^{nN+n^2}.
\end{aligned}
\end{equation}
We have checked it by computing the $x$-series of the both sides for various $N$.
It is quite interesting to prove it rigorously.

To close this subsection, we comment on an equivalence to a previous conjecture proposed in \cite{Hatsuda:2024uwt}.
The formula \eqref{eq:main} is rewritten as
\begin{equation}
\begin{aligned}
I_N(t,u;q)&=\frac{(q;q)_\infty}{(t;t)_{N} (t^Nq;q)_{\infty}} \sum_{n_1=0}^\infty\sum_{n_2=0}^\infty \cdots \sum_{n_N=0}^\infty u^{n_1+2n_2+\cdots N n_N}\\
&\quad \times \frac{(t^{N};q)_{n_1+\cdots+n_N} (t^{N-1};q)_{n_2+\cdots+n_N}\cdots (t;q)_{n_N}}{(t^{N-1}q;q)_{n_1+\cdots+n_N} (t^{N-2}q;q)_{n_2+\cdots+n_N}\cdots (q;q)_{n_N}},
\end{aligned}
\end{equation}
by using \eqref{eq:lambda-n} for $\ell=N$. Using \eqref{eq:q-Poch-id-1} and
\begin{align}
(x;q)_n=\frac{(x;q)_\infty}{(xq^n;q)_\infty},
\end{align}
and imposing $tu=q$, this expression is equivalent to Eq.~(3.55) in \cite{Hatsuda:2024uwt}. In other words, we rigorously derived the earlier conjecture on the flavored Schur index by using known results on the Macdonald polynomials.

\subsection{Comparison with character expansion method}
In the previous subsection, we have shown that the matrix integral of the deformed Schur index is beautifully evaluated with the help of the Macdonald polynomials. There is another powerful tool to evaluate matrix integrals, a.k.a.\ the character expansion method. We compare these two methods.

Let us recall the character expansion method \cite{Dolan:2007rq}. The idea is simple. We start with the matrix integral \eqref{eq:SCI-2}. We denote $f_n=f(t^n,u^n,v^n;p^n,q^n)$ for short, and then the integrand is expanded as
\begin{equation}
\begin{aligned}
\exp \left( \sum_{n=1}^\infty \frac{f_n}{n} p_n(x) p_n(x^{-1}) \right)=\sum_\mu \frac{f_\mu}{z_\mu} p_\mu(x) p_\mu(x^{-1})
\end{aligned}
\end{equation}
where for a partition $\mu=(\mu_1, \mu_2, \dots)=(1^{m_1}2^{m_2}\dots)$, we define
\begin{align}
f_\mu=\prod_{i=1}^{\ell(\mu)} f_{\mu_i}, \qquad p_\mu(x)=\prod_{i=1}^{\ell(\mu)}p_{\mu_i}(x) ,\qquad z_\mu=\prod_{i\geq 1} i^{m_i} m_i! \,.
\end{align}
See Appendix~\ref{app:Mac} for the notation on partitions.
We can translate the power sum symmetric polynomials $p_\mu(x)$ into the Schur polynomials $s_\lambda(x)$ by the so-called Frobenius formula:
\begin{equation}
\begin{aligned}
p_\mu(x)=\sum_{\substack{\lambda \vdash n \\ \ell(\lambda) \leq N}} \chi_{\mu}^{\lambda} s_\lambda(x),
\end{aligned}
\end{equation}
where $n=|\mu|$, and the sum is taken over all the partitions $\lambda$ of $n$ with $\ell(\lambda) \leq N$.  $\chi_{\mu}^{\lambda}$ is the character of the symmetric group $S_n$ for the representation $\lambda$ and the conjugacy class $\mu$. Then we can perform the torus integral by using the orthonormality of the Schur polynomials, and finally obtain
\begin{equation}
\begin{aligned}
I_N(t,u,v;p,q)=\sum_\mu \frac{f_\mu}{z_\mu} \sum_{\substack{\lambda \vdash n \\ \ell(\lambda) \leq N}} (\chi_{\mu}^{\lambda})^2.
\end{aligned}
\label{eq:CEM}
\end{equation}
This character expansion method is powerful to evaluate the superconformal index.
However, for $p=v=0$, our result \eqref{eq:main} has several advantages. The first one is that our formula contains the single partition sum, while \eqref{eq:CEM} has the double partition sums. The computational cost should be saved in our result. The second one is that the values of the character of the symmetric group does not have a general explicit formula. One needs character tables, the Murnaghan-Nakayama rule or the Frobenius formula to obtain them. In our formula, all the ingredients are written in an explicit way. The third and biggest one is that in the character expansion method, it is hard to find exact expressions of the index. 

To see the third point more concretely, let us consider the $1/4$ BPS index: $f(t,u;0)=t+u-tu$ for $N=3$. The character expansion method yields the double series expansion for both $t$ and $u$:
\begin{equation}
\begin{aligned}
I_3(t,u;0)=1+(t+u)+(2t^2+tu+2u^2)+(3t^3+2t^2 u+2t u^2+3u^3)+\cdots
\end{aligned}
\end{equation}
It is quite non-trivial to resum this expansion exactly, and find the analytic form \eqref{eq:quarter}.
Therefore to explore the analytic structure of indices, the character expansion method is not very useful. This point is also important to explore finite $N$ corrections in Section~\ref{sec:GGE}.

\subsection{Line operator indices}\label{sec:line}
We can extend the previous calculation to insertions of characters of $U(N)$.
These are called (Wilson or electric) line operator indices. Let us consider an insertion
\begin{equation}
\begin{aligned}
I_{\lambda, \rho, N}(t,u;q)&=\frac{1}{N!}\frac{(q;q)_\infty^{N}(tu;q)_\infty^N}{(t;q)_\infty^N (u;q)_\infty^N} \oint_{\mathbb{T}^N} \prod_{i=1}^N \frac{dx_i}{2\pi i x_i}\\
&\quad \times \prod_{1\leq i \ne j \leq N} \frac{(x_i/x_j;q)_\infty (t u x_i/x_j;q)_\infty}{(tx_i/x_j;q)_\infty (ux_i/x_j;q)_\infty}
s_{\lambda}(x)s_{\rho}(x^{-1}).
\end{aligned}
\label{eq:LOI}
\end{equation}
Recall that the Schur polynomial $s_\lambda(x)$ is the $U(N)$ character for the representation $\lambda$. We also use a short-hand notation $I_{\lambda, N}(t,u;q)=I_{\lambda,\lambda,N}(t,u;q)$.

We first focus on an anti-symmetric representation $\lambda=\rho=(1^r)$. In this case, the Schur polynomial becomes the elementary symmetric polynomial: $s_{(1^r)}(x)=e_r(x)$. We start with
\begin{equation}
\begin{aligned}
I_{(1^r), N}(t,u;q)&=\frac{1}{N!}\frac{(q;q)_\infty^{N}}{(t;q)_\infty^N} \oint_{\mathbb{T}^N} \prod_{i=1}^N \frac{dx_i}{2\pi i x_i}w(x)\\
&\quad \times  \sum_{\ell(\lambda) \leq N} u^{|\lambda|} b_\lambda P_\lambda(x;q,t)P_\lambda(x^{-1};q,t)e_{r}(x)e_{r}(x^{-1}).
\end{aligned}
\end{equation}
We can use the Pieri formula:
\begin{equation}
\begin{aligned}
e_r(x)P_\lambda(x;q,t)=\sum_{\mu \in V_N^r(\lambda)} \psi_{\mu/\lambda}'(q,t) P_{\mu}(x;q,t),
\end{aligned}
\end{equation}
where $V_N^r(\lambda)$ is defined by \eqref{eq:Vnr}, 
and
\begin{align}
\psi_{\mu/\lambda}'(q,t)&=\psi_{\mu'/\lambda'}(t,q),\\
\psi_{\mu/\lambda}(q,t)&=\prod_{1\leq i \leq j \leq \ell(\lambda)} \frac{(t^{j-i+1}q^{\lambda_i-\lambda_j};q)_{\mu_i-\lambda_i}(t^{j-i}q^{\lambda_i-\mu_{j+1}+1};q)_{\mu_i-\lambda_i}}{(t^{j-i}q^{\lambda_i-\lambda_j+1};q)_{\mu_i-\lambda_i}(t^{j-i+1}q^{\lambda_i-\mu_{j+1}};q)_{\mu_i-\lambda_i}}.
\end{align}
As explained in Appendix~\ref{app:Mac}, $\lambda'$ denotes the conjugate partition of $\lambda$.
With the help of this formula, we can immediately evaluate the torus integral, and obtain
\begin{equation}
\begin{aligned}
I_{(1^r), N}(t,u;q)=\frac{(q;q)_\infty^{N}}{(t;q)_\infty^N}\sum_{\ell(\lambda) \leq N} u^{|\lambda|}b_\lambda \sum_{\mu \in V_N^r(\lambda)} \cN_{\mu,N}\psi_{\mu/\lambda}'(q,t)^2.
\end{aligned}
\end{equation}
We further rewrite it, by using \eqref{eq:bN}, as 
\begin{equation}
\begin{aligned}
I_{(1^r), N}(t,u;q)=\frac{(q;q)}{(t;t)_N (t^N q; q)_\infty}\sum_{\ell(\lambda) \leq N} u^{|\lambda|} \sum_{\mu \in V_N^r(\lambda)} 
\varphi_{\mu/\lambda}'(q,t) \psi_{\mu/\lambda}'(q,t)\\
\times \prod_{i=1}^{\ell(\mu)} \frac{(t^{N-i+1};q)_{\mu_i}}{(t^{N-i}q;q)_{\mu_i}}
\end{aligned}
\label{eq:I-antisym}
\end{equation}
where
\begin{align}
\varphi_{\mu/\lambda}'(q,t)&=\frac{b_{\lambda}}{b_\mu}\psi_{\mu/\lambda}'(q,t)=\varphi_{\mu'/\lambda'}(t,q),\\
\varphi_{\mu/\lambda}(q,t)&=\prod_{1\leq i \leq j \leq \ell(\mu)} \frac{(t^{j-i+1}q^{\mu_i-\mu_j};q)_{\mu_j-\lambda_j}(t^{j-i}q^{\lambda_i-\mu_{j+1}+1};q)_{\mu_{j+1}-\lambda_{j+1}}}{(t^{j-i}q^{\mu_i-\mu_j+1};q)_{\mu_j-\lambda_j}(t^{j-i+1}q^{\lambda_i-\mu_{j+1}};q)_{\mu_{j+1}-\lambda_{j+1}}}.
\end{align}
The combinatorial sums in \eqref{eq:I-antisym} may be implemented in a symbolic computational system.

One way to evaluate the line operator indices for general representations is to use the Jacobi-Trudi formula:
\begin{equation}
\begin{aligned}
s_\lambda(x)=\det(e_{\lambda_i'-i+j}(x))_{1\leq i, j \leq l(\lambda')}.
\end{aligned}
\end{equation}
Then we repeatedly apply the Pieri formula. The resulting formulae are however terribly complicated.
To sketch it, let us consider two-column repsentations $\lambda=(1^{r_1}2^{r_2})$ and $\rho=(1^{r_1'}2^{r_2'})$ in \eqref{eq:LOI}.
Since the Schur polynomial for $\lambda=(1^{r_1}2^{r_2})$ is given by
\begin{equation}
\begin{aligned}
s_{\lambda}(x)&=\begin{vmatrix}
e_{r_1+r_2}(x) & e_{r_1+r_2+1}(x) \\ e_{r_2-1}(x) & e_{r_2}(x)
\end{vmatrix}\\
&=e_{r_1+r_2}(x)e_{r_2}(x)-e_{r_1+r_2+1}(x)e_{r_2-1}(x),
\end{aligned}
\end{equation}
we have
\begin{equation}
\begin{aligned}
&s_{\lambda}(x)s_{\rho}(x^{-1})\\
&=e_{r_1+r_2}(x)e_{r_2}(x)e_{r_1'+r_2'}(x^{-1})e_{r_2'}(x^{-1})-e_{r_1+r_2}(x)e_{r_2}(x)e_{r_1'+r_2'+1}(x^{-1})e_{r_2'-1}(x^{-1})\\
&\quad -e_{r_1+r_2+1}(x)e_{r_2-1}(x)e_{r_1'+r_2'}(x^{-1})e_{r_2'}(x^{-1})\\
&\quad+e_{r_1+r_2+1}(x)e_{r_2-1}(x)e_{r_1'+r_2'+1}(x^{-1})e_{r_2'-1}(x^{-1}).
\end{aligned}
\end{equation}
Since $e_r(x)e_s(x)P_\mu(x;q,t)$ is expanded as $P_\nu(x;q,t)$ by using the Pieri formula twice, we can evaluate the integral of $I_{\lambda, \rho, N}(t,u;q)$. In the next section, we see that in the large $N$ limit, the computation is drastically simplified.

\section{Large $N$ limit}\label{sec:large-N}
\subsection{Analysis at infinite $N$}\label{sec:infinite-N}
In the context of the AdS/CFT correspondence, we are usually interested in the large $N$ limit and finite $N$ corrections to it.
It is not obvious to take the limit $N \to \infty$ in the matrix integral \eqref{eq:SCI} or \eqref{eq:SCI-2}.
One standard way to do so is to use the technique of random matrices, \textit{i.e.}, to use the saddle-point analysis of matrix integrals.
In this subsection, we develop another way to treat the strictly infinite $N$ analysis, based on the theory of symmetric functions.

In combinatorics, it is often useful to consider ``symmetric polynomials with an infinite number of variables''. Such are usually referred to as symmetric functions. The basic philosophy of the famous book \cite{Macdonald} is to develop the theory of symmetric functions rather than symmetric polynomials. Many results on symmetric polynomials of $x=(x_1,\dots, x_N)$ are obtained from those on symmetric functions by projection $x_{N+1}=x_{N+2}=\dots=0$.
Inverting the logic, we easily obtain results at $N=\infty$ in an algebraic way. Note that most of the results in this section are quoted from Chapter VI in \cite{Macdonald}.

Let us define an expectation value and an inner product by
\begin{align}
\vev{A(x)}_N'&=\frac{1}{N!} \oint_{\mathbb{T}^N} \prod_{i=1}^N \frac{dx_i}{2\pi i x_i} w(x) A(x),\\
\langle f, g \rangle_N'&=\frac{1}{N!} \oint_{\mathbb{T}^N} \prod_{i=1}^N \frac{dx_i}{2\pi i x_i} w(x) f(x)g(x^{-1})=\vev{f(x)g(x^{-1})}_N'
\label{eq:inner-N}
\end{align}
In this notation, the deformed Schur index and the half-index are written as
\begin{align}
I_N(t,u;q)&=\frac{(q;q)_\infty^N}{(t,q)_\infty^N} \biggl\langle \prod_{i,j=1}^N \frac{(tu x_i/x_j;q)_\infty}{(ux_i/x_j;q)_\infty} 
\biggr\rangle_N',\\
I_N(t,0;q)&=\frac{(q;q)_\infty^N}{(t,q)_\infty^N}\vev{1}_N'.
\end{align}
Now we define the expectation value and the inner product at $N=\infty$ by
\begin{align}
\vev{A(x)}_\infty=\lim_{N \to \infty} \frac{\vev{A(x)}_N'}{\vev{1}_N'},\qquad
\langle f, g \rangle_\infty=\lim_{N \to \infty} \frac{\langle f, g \rangle_N'}{\langle 1, 1 \rangle_N'}.
\end{align}
In these expressions, the functions on the left hand sides have an infinite number of variables.
The inner product $\langle f, g \rangle_\infty$ has a very nice property. The power sum symmetric functions now satisfy an orthogonal relation,
\begin{equation}
\begin{aligned}
\langle p_\lambda, p_\mu \rangle_\infty=\delta_{\lambda, \mu}z_\lambda(q,t),
\end{aligned}
\end{equation}
where
\begin{equation}
\begin{aligned}
z_\lambda(q,t)=z_\lambda \prod_{i=1}^{\ell(\lambda)} \frac{1-q^{\lambda_i}}{1-t^{\lambda_i}}
\end{aligned}
\end{equation}
Of course, this is not the case for finite $N$. The power sum symmetric polynomials are not orthogonal for the inner product \eqref{eq:inner-N}.

Let us consider a ratio
\begin{equation}
\begin{aligned}
\frac{I_{\infty}(t,u;q)}{I_{\infty}(t,0;q)}=\lim_{N \to \infty} \frac{I_{N}(t,u;q)}{I_{N}(t,0;q)}=\biggl\langle \prod_{i,j=1}^\infty \frac{(tu x_i/x_j;q)_\infty}{(ux_i/x_j;q)_\infty} \biggr\rangle_\infty.
\end{aligned}
\end{equation}
We can easily evaluate it. Using
\begin{equation}
\begin{aligned}
\prod_{i,j=1}^\infty \frac{(tu x_i/x_j;q)_\infty}{(ux_i/x_j;q)_\infty} &=\exp\biggl( \sum_{n=1}^\infty \frac{u^n}{n} \frac{1-t^n}{1-q^n}p_n(x)p_n(x^{-1}) \biggr) \\
&=\sum_{\lambda} \frac{u^{|\lambda|}}{z_\lambda(q,t)} p_\lambda(x)p_\lambda(x^{-1}),
\end{aligned}
\end{equation}
we find
\begin{equation}
\begin{aligned}
\frac{I_{\infty}(t,u;q)}{I_{\infty}(t,0;q)}=\sum_{\lambda} \frac{u^{|\lambda|}}{z_\lambda(q,t)} \langle p_\lambda, p_\lambda \rangle_{\infty}
=\sum_\lambda u^{|\lambda|}=\frac{1}{(u;u)_\infty}.
\end{aligned}
\end{equation}
Since the large $N$ limit of the half-index is easily obtained from the exact result \eqref{eq:half-I} as
\begin{equation}
\begin{aligned}
I_{\infty}(t,0;q)=\frac{(q;q)_\infty}{(t;t)_\infty},
\end{aligned}
\end{equation}
we find
\begin{equation}
\begin{aligned}
I_{\infty}(t,u;q)=\frac{(q;q)_\infty}{(t;t)_\infty (u;u)_\infty}.
\end{aligned}
\end{equation}
We can also re-derive the same result from our exact formula \eqref{eq:main}.
However, we should be careful when taking the large $N$ limit.
First, $b_\lambda$ is written as\footnote{See Eq. (4.11) in \cite{Macdonald}.}
\begin{equation}
\begin{aligned}
b_\lambda=\frac{1}{\vev{P_\lambda, P_\lambda}_\infty}=\lim_{N \to \infty} \frac{\cN_{\emptyset, N}}{\cN_{\lambda, N}}.
\end{aligned}
\end{equation}
Using \eqref{eq:bN}, we find
\begin{equation}
\begin{aligned}
\lim_{N \to \infty} \prod_{i=1}^{\ell(\lambda)} \frac{(t^{N-i+1};q)_{\lambda_i}}{(t^{N-i}q;q)_{\lambda_i}}
=\lim_{N \to \infty} \frac{b_\lambda \cN_{\lambda, N}}{b_{\emptyset} \cN_{\emptyset,N}}=1.
\end{aligned}
\end{equation}
Therefore, from \eqref{eq:main}, we obtain
\begin{equation}
\begin{aligned}
I_\infty(t,u;q)=\frac{(q;q)_\infty}{(t;t)_\infty} \sum_{\lambda} u^{|\lambda|}=\frac{(q;q)_\infty}{(t;t)_\infty (u;u)_\infty}.
\end{aligned}
\end{equation}

To see line operator indices, let us start with an insertion of the power sum symmetric polynomials,
\begin{equation}
\begin{aligned}
I_{\mu, \nu, N}^\text{p.s.}(t,u;q)&=\frac{1}{N!}\frac{(q;q)_\infty^{N}(tu;q)_\infty^N}{(t;q)_\infty^N (u;q)_\infty^N} \oint_{\mathbb{T}^N} \prod_{i=1}^N \frac{dx_i}{2\pi i x_i}\\
&\quad \times \prod_{1\leq i \ne j \leq N} \frac{(x_i/x_j;q)_\infty (t u x_i/x_j;q)_\infty}{(tx_i/x_j;q)_\infty (ux_i/x_j;q)_\infty}
p_{\mu}(x)p_{\nu}(x^{-1}).
\end{aligned}
\end{equation}
We would like to know $I_{\mu, \nu, \infty}^\text{p.s.}(t,u;q)$. To do so, we start with
\begin{equation}
\begin{aligned}
\frac{I_{\mu, \nu, \infty}^\text{p.s.}(t,u;q)}{I_\infty(t,0;q)}=\biggl\langle \prod_{i,j=1}^\infty \frac{(tu x_i/x_j;q)_\infty}{(ux_i/x_j;q)_\infty} p_\mu(x) p_\nu(x^{-1})\biggr\rangle_\infty .
\end{aligned}
\end{equation}
We can still evaluate it as follows. Considering
\begin{equation}
\begin{aligned}
\prod_{i,j=1}^\infty \frac{(tu x_i/x_j;q)_\infty}{(ux_i/x_j;q)_\infty} p_\mu(x) p_\nu(x^{-1})
&=\sum_{\lambda} \frac{u^{|\lambda|}}{z_\lambda(q,t)} p_\lambda(x)p_\lambda(x^{-1})p_\mu(x) p_\nu(x^{-1}) \\
&=\sum_{\lambda} \frac{u^{|\lambda|}}{z_\lambda(q,t)} p_{\lambda \cup \mu}(x)p_{\lambda\cup \nu}(x^{-1}),
\end{aligned}
\end{equation}
where $\lambda \cup \mu$ is a union of two partitions $\lambda$ and $\mu$,%
\footnote{For example, if $\lambda=(3,1)$ and $\mu=(2,2,1)$, then $\lambda \cup \mu=(3,2,2,1,1)$.}
we find
\begin{equation}
\begin{aligned}
\frac{I_{\mu, \nu, \infty}^\text{p.s.}(t,u;q)}{I_\infty(t,0;q)}
&=\sum_{\lambda} \frac{u^{|\lambda|}}{z_\lambda(q,t)} \langle p_{\lambda \cup \mu}, p_{\lambda \cup \nu} \rangle_\infty \\
&= \delta_{\mu, \nu} \sum_{\lambda} \frac{u^{|\lambda|}}{z_\lambda(q,t)}z_{\lambda \cup \mu}(q,t). 
\end{aligned}
\end{equation}
For $\lambda=(1^{k_1}2^{k_2}\dots)$ and $\mu=(1^{m_1}2^{m_2}\dots)$, we have $\lambda \cup \mu=(1^{k_1+m_1}2^{k_2+m_2}\dots)$ and
 \begin{equation}
\begin{aligned}
z_{\lambda \cup \mu}(q,t)=z_\lambda(q,t)z_\mu(q,t)\prod_{i \geq 1} \binom{k_i+m_i}{k_i}.
\end{aligned}
\end{equation}
Then, the sum over $\lambda$ is performed,
\begin{equation}
\begin{aligned}
\frac{I_{\mu, \nu, \infty}^\text{p.s.}(t,u;q)}{I_\infty(t,0;q)}
&=\delta_{\mu, \nu} z_\mu(q,t) \sum_{\lambda}\prod_{i\geq 1} \binom{k_i+m_i}{k_i} u^{ik_i}\\
&=\delta_{\mu, \nu}z_\mu(q,t) \prod_{i\geq 1} \frac{1}{(1-u^i)^{m_i+1}}\\
&=\delta_{\mu, \nu}z_\mu(q,t) \frac{1}{(u;u)_\infty} \prod_{i=1}^{\ell(\mu)} \frac{1}{1-u^{\mu_i}}\\
&=\delta_{\mu, \nu}\frac{z_\mu}{(u;u)_\infty}\prod_{i=1}^{\ell(\mu)} \frac{1-q^{\mu_i}}{(1-t^{\mu_i})(1-u^{\mu_i})}.
\end{aligned}
\end{equation}
Therefore
\begin{equation}
\begin{aligned}
I_{\mu, \nu, \infty}^\text{p.s.}(t,u;q)=\frac{(q;q)_\infty}{(t;t)_\infty (u;u)_\infty} \delta_{\mu, \nu}z_\mu\prod_{i=1}^{\ell(\mu)}\frac{1-q^{\mu_i}}{(1-t^{\mu_i})(1-u^{\mu_i})}.
\end{aligned}
\end{equation}
The same result was obtained in \cite{Hatsuda:2023iwi, Hatsuda:2023imp} by the Fermi-gas formalism and in \cite{Imamura:2024zvw} by the character expansion method.
Using the Frobenius formula:
\begin{equation}
\begin{aligned}
s_\lambda(x)=\sum_{\mu \vdash |\lambda|} \frac{\chi_\mu^\lambda}{z_\mu} p_\mu(x),
\end{aligned}
\end{equation}
we finally arrive at the general line operator index at $N=\infty$,
\begin{equation}
\begin{aligned}
\frac{I_{\lambda,\rho,\infty}(t,u;q)}{I_\infty(t,u;q)}
=\sum_{\mu \vdash |\lambda|} \frac{\chi_\mu^\lambda \chi_\mu^\rho}{z_\mu} \prod_{i=1}^{\ell(\mu)}\frac{1-q^{\mu_i}}{(1-t^{\mu_i})(1-u^{\mu_i})}.
\end{aligned}
\end{equation}
If $\lambda=\rho=(1^r)$ or $\lambda=\rho=(r)$, we have $\chi_\mu^\lambda \chi_\mu^\rho=1$ for any $\mu$. We then find
\begin{equation}
\begin{aligned}
\frac{I_{(1^r),\infty}(t,u;q)}{I_\infty(t,u;q)}
=\frac{I_{(r),\infty}(t,u;q)}{I_\infty(t,u;q)}
=\sum_{\mu \vdash r} \frac{1}{z_\mu} \prod_{i=1}^{\ell(\mu)}\frac{1-q^{\mu_i}}{(1-t^{\mu_i})(1-u^{\mu_i})}.
\end{aligned}
\end{equation}
If taking $N \to \infty$ in \eqref{eq:I-antisym}, we obtain
\begin{equation}
\begin{aligned}
\frac{I_{(1^r),\infty}(t,u;q)}{I_\infty(t,u;q)}=(u;u)_\infty \sum_\lambda u^{|\lambda|} \sum_{\mu \in V_{\infty}^r(\lambda)} \varphi'_{\mu/\lambda}(q,t) \psi'_{\mu/\lambda}(q,t).
\end{aligned}
\end{equation}
These two must be equal.

\subsection{Finite $N$ corrections: giant graviton expansions}\label{sec:GGE}
One of the most remarkable properties on superconformal indices is that their finite $N$ corrections are also generated by  (analytically continued) superconformal indices \cite{Arai:2019xmp, Arai:2020qaj, Imamura:2021ytr, Gaiotto:2021xce} (see also \cite{Murthy:2022ien, Beccaria:2023zjw, Beccaria:2023hip}). From a perspective of the AdS/CFT correspondence, this property is often referred to as a giant graviton expansion. In this section, we study such a surprising structure, particularly found by Gaiotto and Lee in \cite{Gaiotto:2021xce} because this type of expansion is well suited for our formula \eqref{eq:main}.

Their basic claim is that the finite $N$ corrections to the superconformal index is given by
\begin{equation}
\begin{aligned}
\frac{I_N(t,u,v;p,q)}{I_\infty(t,u,v;p,q)}=\sum_{k=0}^\infty t^{kN} \hat{I}_k(t,u,v;p,q),
\end{aligned}
\end{equation}
where $\hat{I}_k(t,u,v;p,q)$ is another index for gauge group $U(k)$, whose single letter index $\hat{f}(t,u,v;p,q)$ is determined by the condition:
\begin{equation}
\begin{aligned}
(1-f)(1-\hat{f})=(1-t)(1-t^{-1}).
\end{aligned}
\end{equation}
It is very easy to see that $\hat{f}(t,u,v;p,q)$ is given by
\begin{equation}
\begin{aligned}
\hat{f}(t,u,v;p,q)=1-\frac{(1-t^{-1})(1-p)(1-q)}{(1-u)(1-v)}.
\end{aligned}
\end{equation}
This means 
\begin{equation}
\begin{aligned}
\hat{I}_k(t,u,v;p,q)=I_k(t^{-1},q,p;v,u).
\end{aligned}
\end{equation}
Note that from \eqref{eq:constraint} we have $t^{-1}pq=uv$. 
To the author's knowledge, this proposal is yet to be proved, but has been confirmed in various limits.

In our interested case $v=p=0$, we have $\hat{I}_k(t,u,0;0,q)=I_k(t^{-1},q,0;0,u)$.
Therefore the giant graviton expansion for the deformed Schur index is given by
\begin{equation}
\begin{aligned}
\frac{I_N(t,u;q)}{I_\infty(t,u;q)}=\sum_{k=0}^\infty t^{kN} \hat{I}_k(t,u;q),\qquad
\hat{I}_k(t,u;q)=I_k(t^{-1},q;u),
\end{aligned}
\label{eq:GGE}
\end{equation}
where $\hat{I}_0(t,u;q):=1$. There are two subtleties to check this highly non-trivial claim.

One is that the ``giant graviton index'' $\hat{I}_k(t,u;q)=I_k(t^{-1},q;u)$ should be understood as an analytic continuation of the original index because the first fugacity $t^{-1}$ satisfies $|t^{-1}|>1$ when $|t|<1$, which is a condition for the convergence of the matrix integral of the original index.
This problem is not a problem in our formula \eqref{eq:main} because it is exact in $t$. We can analytically continue it to $|t|>1$ regime.
Note that in the character expansion method, one has to resum the power series of $t$ in \eqref{eq:CEM} for the analytic continuation. This resummation is non-trivial.
For the 1/4 BPS index $I_N(t,u;0)$, the giant graviton index $\hat{I}_k(t,u;0)=I_k(t^{-1},0;u)$ is the analytic continuation of the half-index.
For the half-index $I_N(t,0;q)$, the giant graviton index $\hat{I}_k(t,0;q)=I_k(t^{-1},q;0)$ is conversely the analytic continuation of the 1/4 BPS index. In these cases, we can prove the giant graviton expansions analytically \cite{Hatsuda:2024uwt}.

The other is the exchange between $u$ and $q$. Our formula \eqref{eq:main} is given by a power series of $u$ but exact in $t$ and $q$. On the other hand, $I_k(t^{-1},q;u)$ has a power series of $q$, not $u$. To resolve this mismatch of the expansion regimes, we scale both $u$ and $q$ simultaneously. For example, we set
\begin{equation}
\begin{aligned}
q=\alpha u.
\end{aligned}
\end{equation}
In the following analysis, we consider this parametrization. Note that the flavored Schur index corresponds to $\alpha=t$.

We follow the argument in \cite{Gaiotto:2021xce}.
We first expand $I_N(t,u;\alpha u)$ with respect to $u$. Using our formula \eqref{eq:main}, we find
\begin{align}
I_1(t,u;\alpha u)&=\frac{1}{1-t}+(1-\alpha)u+(1-\alpha)(1+\alpha+\alpha t)u^2+O(u^3), \\
I_2(t,u;\alpha u)&=\frac{1}{(1-t)(1-t^2)}+\frac{1-\alpha}{1-t}u+\frac{(1-\alpha)(2+\alpha-t+\alpha t^2)}{1-t}u^2+O(u^3), \\
I_3(t,u;\alpha u)&=\frac{1}{(1-t)(1-t^2)(1-t^3)}+\frac{1-\alpha}{(1-t)(1-t^2)}u \notag\\
&\hspace{4truecm}+\frac{(1-\alpha)(2+\alpha-t^2+\alpha t^3)}{(1-t)(1-t^2)}u^2+O(u^3).
\end{align}
In general, $I_N(t,u;\alpha u)$ has the following nice structure:
\begin{equation}
\begin{aligned}
I_N(t,u;\alpha u)=\sum_{j=0}^\infty g_{N,j}^{(\alpha)}(t) u^j,\qquad
g_{N,j}^{(\alpha)}(t)=\frac{1-\alpha}{(t;t)_{N-1}}G_{N,j}^{(\alpha)}(t),
\end{aligned}
\label{eq:u-series}
\end{equation}
where the explicit forms of $G_{N,j}^{(\alpha)}(t)$ for $j=0,1,2$ are given by
\begin{align}
G_{N,0}^{(\alpha)}(t)&=\frac{1}{(1-\alpha)(1-t^{N})}, \\
G_{N,1}^{(\alpha)}(t)&=1, \\
G_{N,2}^{(\alpha)}(t)&=2+\alpha+t^N\biggl( -\frac{1}{t}+\alpha \biggr).
\label{eq:GN2}
\end{align}
We observe that $G_{N,j}^{(\alpha)}(t)$ for $j \geq 1$ is a ``polynomial'' of degree $j-1$ in $t^N$.

If we assume the giant graviton expansion \eqref{eq:GGE}, we can fix $G_{N,j}^{(\alpha)}(t)$ recursively.
Let us introduce
\begin{equation}
\begin{aligned}
I_\infty(t,u;\alpha u)=\frac{(\alpha u;\alpha u)_\infty}{(t;t)_\infty (u;u)_\infty}=\sum_{j=0}^\infty g_{\infty,j}^{(\alpha)}(t) u^j,\qquad
g_{\infty,j}^{(\alpha)}(t)=\frac{1-\alpha}{(t;t)_{\infty}}G_{\infty,j}^{(\alpha)}.
\end{aligned}
\label{eq:u-series-inf}
\end{equation}
Note that $I_k(t^{-1},\alpha u;u)$ has the following expansion:
\begin{equation}
\begin{aligned}
I_k(t^{-1},\alpha u;u)=I_k(t^{-1}, \alpha u; \alpha^{-1}\cdot \alpha u)=\sum_{j=0}^\infty g_{k,j}^{(\alpha^{-1})}(t^{-1})\alpha^j u^j.
\end{aligned}
\label{eq:u-series-dual}
\end{equation}
Plugging \eqref{eq:u-series}, \eqref{eq:u-series-inf} and \eqref{eq:u-series-dual} into \eqref{eq:GGE}, we obtain
\begin{equation}
\begin{aligned}
g_{N,j}^{(\alpha)}(t)=\sum_{k=0}^\infty t^{kN} \sum_{m=0}^j \alpha^{m} g_{\infty, j-m}^{(\alpha)}(t) g_{k,m}^{(\alpha^{-1})}(t^{-1}) \qquad(j \geq 1) ,
\end{aligned}
\label{eq:rel-for-g}
\end{equation}
where $g_{0,m}^{(\alpha^{-1})}(t^{-1})=1$.

We further translate \eqref{eq:rel-for-g} into that for $G_{N,j}^{(\alpha)}(t)$ and assume that $G_{N,j}^{(\alpha)}(t)$ is a polynomial of degree $j-1$ in $t^N$.
We finally obtain
\begin{equation}
\begin{aligned}
G_{N,j}^{(\alpha)}(t)&=G_{\infty, j}^{(\alpha)}+\sum_{n=1}^{j-1} t^{nN}\biggl[ \frac{G_{\infty, j}^{(\alpha)}}{(t;t)_n}\\
&\quad+\sum_{k=1}^n \sum_{m=0}^j \frac{(-1)^{k-1}t^{\frac{k(k-1)}{2}}\alpha^m(1-\alpha^{-1})}{(t;t)_{n-k}(t;t)_{k-1}}G_{\infty, j-m}^{(\alpha)}G_{k,m}^{(\alpha^{-1})}(t^{-1})\biggr] \quad (j \geq 1).
\end{aligned}
\end{equation}
To fix $G_{N,j}^{(\alpha)}(t)$ recursively from this relation, we need inputs $G_{k,j}^{(\alpha^{-1})}(t^{-1})$ for $k=1,2,\dots j-1$. For low values of $j$, this is easily done by using \eqref{eq:main}. For example, for $j=2$, we need only $G_{1,2}^{(\alpha^{-1})}(t^{-1})$:
\begin{equation}
\begin{aligned}
G_{N,2}^{(\alpha)}(t)=2+\alpha+t^N\bigl(1+2\alpha-\alpha G_{1,2}^{(\alpha^{-1})}(t^{-1}) \bigr).
\end{aligned}
\end{equation}
Using $G_{1,2}^{(\alpha)}(t)=1+\alpha+\alpha t$, this reproduces the previous result \eqref{eq:GN2}.
Pushing this computation, we further find
\begin{equation}
\begin{aligned}
G_{N,3}^{(\alpha)}(t)&=3+\alpha+t^N\biggl(-\frac{1}{t^2}+\frac{-2+\alpha}{t}+\alpha+\alpha^2\biggr)+t^{2N}\biggl(\frac{1}{t^3}-\frac{\alpha}{t^2}-\frac{\alpha}{t}+\alpha^2 \biggr),\\
G_{N,4}^{(\alpha)}(t)&=5+2\alpha+t^N\biggl(-\frac{1}{t^3}+\frac{-2+\alpha}{t^2}+\frac{-4+\alpha+\alpha^2}{t}+\alpha+2\alpha^2+\alpha^3\biggr)\\
&\quad+t^{2N}\biggl(\frac{1}{t^5}+\frac{1-\alpha}{t^4}+\frac{2-2\alpha}{t^3}+\frac{-3\alpha+\alpha^2}{t^2}-\frac{\alpha}{t}+\alpha^2+\alpha^3\biggr)\\
&\quad+t^{3N}\biggl(-\frac{1}{t^6}+\frac{\alpha}{t^5}+\frac{\alpha}{t^4}+\frac{\alpha-\alpha^2}{t^3}-\frac{\alpha^2}{t^2}-\frac{\alpha^2}{t}+\alpha^3\biggr).
\end{aligned}
\end{equation}
The high-$j$ computations are straightforward.
It would be interesting to find the general structure of $G_{N,j}^{(\alpha)}(t)$.

We can repeat the same computation for line operator indices. 
The giant graviton expansions (or brane expansions) of the line operator indices were studied in \cite{Imamura:2024lkw, Beccaria:2024oif, Imamura:2024pgp, Imamura:2024zvw}.
Here we focus on the fundamental representation. From \eqref{eq:I-antisym}, we have
\begin{align}
I_{(1), 1}(t,u;\alpha u)&=\frac{1}{1-t}+(1-\alpha)u+(1-\alpha)(1+\alpha+\alpha t)u^2+O(u^3), \\
I_{(1), 2}(t,u;\alpha u)&=\frac{1}{(1-t)^2}+\frac{1-\alpha}{1-t}(2+t)u \notag \\
&\quad+\frac{1-\alpha}{1-t}\Bigl(3+2\alpha t+(-1+2\alpha)t^2+\alpha t^3\Bigr)u^2+O(u^3),\\
I_{(1), 3}(t,u;\alpha u)&=\frac{1}{(1-t)^2(1-t^2)}+\frac{1-\alpha}{(1-t)(1-t^2)}(2+2t+t^2)u\notag \\
&\quad+\frac{1-\alpha}{(1-t)(1-t^2)}\Bigl( 4+3t+(-1+2\alpha)t^2+(-2+3\alpha)t^3\notag \\
&\quad+(-1+2\alpha)t^4+\alpha t^5\Bigr)u^2+O(u^3).
\end{align}
There is the following nice structure:
\begin{equation}
\begin{aligned}
I_{(1), N}(t,u;\alpha u)=\frac{1-\alpha}{(1-t)(t;t)_{N-1}}\sum_{j=0}^\infty u^j G_{(1), N, j}^{(\alpha)}(t),
\end{aligned}
\label{eq:I_fund}
\end{equation}
where
\begin{equation}
\begin{aligned}
G_{(1), N, 0}^{(\alpha)}(t)=\frac{1}{1-\alpha}.
\end{aligned}
\end{equation}
We would like to determine $G_{(1), N, j}^{(\alpha)}(t)$ for $j\geq 1$ from the giant graviton expansion.
The giant graviton expansion for the fundamental line operator index was proposed in \cite{Imamura:2024lkw},
\begin{equation}
\begin{aligned}
\frac{I_{(1), N}(t,u;q)}{I_{(1), \infty}(t,u;q)}
=\frac{I_{N}(t,u;q)}{I_{\infty}(t,u;q)}-\frac{(1-t^{-1})(1-q)}{1-u} \sum_{k=1}^\infty t^{kN} I_{(1), k}(t^{-1}, q; u),
\end{aligned}
\label{eq:I_fund-GGE}
\end{equation}
where
\begin{equation}
\begin{aligned}
I_{(1), \infty}(t,u;q)=\frac{1-q}{(1-t)(1-u)}I_{\infty}(t,u;q).
\end{aligned}
\end{equation}
Plugging the ansatz \eqref{eq:I_fund} into \eqref{eq:I_fund-GGE}, we find an analytic form of $G_{(1), N, j}^{(\alpha)}(t)$ ($j=1,2,3$),
\begin{align}
G_{(1), N, 1}^{(\alpha)}(t)&=2-t^{N}\biggl(\frac{1}{t}+1\biggr),\\
G_{(1), N, 2}^{(\alpha)}(t)&=4+t^N\biggl( -\frac{1}{t^2}+\frac{2(-2+\alpha)}{t}-2+\alpha \biggr)\notag \\
&\quad+t^{2N}\biggl(\frac{1}{t^3}+\frac{1-\alpha}{t^2}+\frac{1-\alpha}{t}-\alpha \biggr),\\
G_{(1), N, 3}^{(\alpha)}(t)&=7-\alpha-\alpha^2+t^N\biggl( -\frac{1}{t^3}+\frac{-4+3\alpha}{t^2}+\frac{-9+7\alpha}{t}-3+\alpha+\alpha^2 \biggr)\notag \\
&\hspace{-1truecm}+t^{2N}\biggl(\frac{1}{t^5}+\frac{2-\alpha}{t^4}+\frac{5(1-\alpha)}{t^3}+\frac{4-7\alpha+2\alpha^2}{t^2}+\frac{2-4\alpha+\alpha^2}{t}-\alpha+\alpha^2 \biggr)\notag\\
&\hspace{-1truecm}+t^{3N}\biggl(-\frac{1}{t^6}+\frac{-1+\alpha}{t^5}+\frac{-1+2\alpha}{t^4}-\frac{(1-\alpha)^2}{t^3}+\frac{(2-\alpha)\alpha}{t^2}+\frac{(1-\alpha)\alpha}{t}-\alpha^2 \biggr).
\end{align}

\section{Conclusion}\label{sec:conclusion}
In this work, we found a new technique to evaluate the unitary matrix integral in the two-parameter deformation of the Schur index. We used known mathematical results on the Macdonald polynomials. The resulting expression \eqref{eq:main} is quite simple, and it is particularly useful in the study of finite $N$ corrections.

There are several directions to future works. It is desirable to extend the formalism in this work to the full superconformal index \eqref{eq:SCI}. It is known that the matrix integral \eqref{eq:SCI} can be rewritten in terms of the elliptic gamma function \cite{Dolan:2008qi}, which is a one-parameter deformation of the $q$-Pochhammer symbol. To evaluate the matrix integral \eqref{eq:SCI} along this line, we probably need an ``elliptic deformation'' of the Macdonald polynomials. It would be interesting to develop it.

It is also intriguing to explore the S-duality between Wilson line operators and 't Hooft line operators. For instance, it is known that the Wilson line operator index \eqref{eq:I-antisym} for the anti-symmetric representation has the S-dual description by the 't Hooft line operator. According to \cite{Gang:2012yr}, the corresponding 't Hooft line operator index in the flavored Schur limit is given by
\begin{equation}
\begin{aligned}
I_{(1^r, 0^{N-r})}^\text{'t Hooft}&=\frac{1}{r!(N-r)!}\frac{(q;q)_\infty^{2N}}{(t;q)_\infty^N (u;q)_\infty^N}\oint_{\mathbb{T}^r} \prod_{i=1}^r \frac{dx_i}{2\pi i x_i} \oint_{\mathbb{T}^{N-r}}\prod_{j=1}^{N-r} \frac{dy_j}{2\pi i y_j}\\
& \times \prod_{1 \leq i \ne j \leq r} \frac{(x_i/x_j;q)_\infty (q x_i/x_j;q)_\infty}{(tx_i/x_j;q)_\infty (ux_i/x_j;q)_\infty}
\prod_{1 \leq i \ne j \leq N-r} \frac{(y_i/y_j;q)_\infty (q y_i/y_j;q)_\infty}{(ty_i/y_j;q)_\infty (uy_i/y_j;q)_\infty} \\
&\hspace{-0.4truecm} \times \prod_{i=1}^r \prod_{j=1}^{N-r} \frac{(q^{1/2}x_i/y_j;q)_\infty (q^{1/2}y_j/x_i;q)_\infty (q^{3/2}x_i/y_j;q)_\infty (q^{3/2}y_j/x_i;q)_\infty}{(tq^{1/2}x_i/y_j;q)_\infty (tq^{1/2}y_j/x_i;q)_\infty (uq^{1/2}x_i/y_j;q)_\infty (uq^{1/2}y_j/x_i;q)_\infty},
\end{aligned}
\end{equation}
where $q=tu$. So far, we do not have a nice way to evaluate this integral exactly. It would be interesting to find it and to prove the equivalence.

An extension to other gauge groups is also interesting.  The Schur (line operator) indices and their giant graviton expansions of type BCD are extensively studied in \cite{Spiridonov:2010qv, Sei:2023fjk, Du:2023kfu, Fujiwara:2023bdc, Hatsuda:2024lcc, Hatsuda:2025jze}. For general root systems, the Macdonald polynomials can be also defined \cite{Macdonald_2003}. They are unified by the so-called Koornwinder polynomials \cite{Koornwinder1991AskeyWilsonPF}. It would be nice to use the Macdonald-Koornwinder polynomials to evaluate the deformed Schur indices for general root systems.

\acknowledgments{
I thank Tadashi Okazaki and Shintarou Yanagida for useful discussions. I'm especially very grateful to Masatoshi Noumi for telling me much about the Macdonald-Koornwinder polynomials in Rikkyo University. This project began with my realization while reading Noumi-san's book \cite{Noumi}, which is quite ``physicist-friendly''. This work was supported in part by JSPS KAKENHI Grant Nos. 22K03641 and
23K25790.
}

\appendix

\section{Review of Macdonald polynomials}\label{app:Mac}
In this appendix, we quickly review Macdonald polynomials of type A. We basically follow the notation in \cite{Macdonald, Noumi}.

\paragraph{Partitions.}
Let $\lambda$ be a partition. We denote it by
\begin{equation}
\begin{aligned}
\lambda=(\lambda_1,\lambda_2,\dots),\qquad \lambda_1 \geq \lambda_2 \geq \dots \geq 0,
\end{aligned}
\end{equation}
or by
\begin{equation}
\begin{aligned}
\lambda=(1^{m_1}2^{m_2}\dots),\qquad m_i \geq 0.
\end{aligned}
\end{equation}
Here $\lambda_i$ are called parts of $\lambda$, and $m_i$ multiplicity of $i$.
The number of non-zero parts $\lambda_i$ is called length, denoted by $\ell(\lambda)$.
The weight $|\lambda|$ is the sum of the parts,
\begin{equation}
\begin{aligned}
|\lambda|=\lambda_1+\lambda_2+\cdots=m_1+2m_2+\cdots.
\end{aligned}
\end{equation}
If $|\lambda|=n$, then $\lambda$ is called a partition of $n$. We denote it by $\lambda \vdash n$.
A partition has a one-to-one correspondence to a Young diagram.
We sometimes identify a partition with its corresponding Young diagram. 
Let us consider a Young diagram for $\lambda$. The partition for the transposed Young diagram is called the conjugate partition of $\lambda$, which is denoted by $\lambda'$. For example, if $\lambda=(7,5,4,1)$, then $\ell(\lambda)=4$, $|\lambda|=17$ and $\lambda'=(4,3,3,3,2,1,1)$.

Let $\lambda$ and $\mu$ be two partitions. If $\lambda_i \geq \mu_i$ for any $i=1,2,\dots$, we denote $\lambda \supset \mu$. In this case, the Young diagram for $\lambda$ includes that for $\mu$. We can subtract the diagram $\mu$ from $\lambda$. The remaining one is referred to as a skew diagram, denoted by $\lambda/\mu$. If two partitions $\lambda$ and $\mu$ satisfy
\begin{equation}
\begin{aligned}
\lambda_1 \geq \mu_1 \geq \lambda_2 \geq \mu_2 \geq \lambda_3 \geq \dots ,
\end{aligned}
\end{equation}
then the skew diagram $\lambda/\mu$ is called a horizontal strip. It also satisfies $\lambda_i'-\mu_i' \leq 1$ for any $i=1,2,\dots$.
Similarly, if $\lambda$ and $\mu$ satisfy
\begin{equation}
\begin{aligned}
\lambda_1' \geq \mu_1' \geq \lambda_2' \geq \mu_2' \geq \lambda_3' \geq \dots ,
\end{aligned}
\end{equation}
then the skew diagram $\lambda/\mu$ is called a vertical strip. For the vertical strip, $\lambda_i-\mu_i \leq 1$ holds for any $i=1,2,\dots$. For example, if $\lambda=(3,3,1)$ and $\mu=(3,1)$, then $\lambda/\mu$ is a horizontal strip, but not a vertical strip.

We also introduce the dominance ordering of partitions:
\begin{equation}
\begin{aligned}
\mu \leq \lambda \quad \Longleftrightarrow \quad |\mu|=|\lambda| \quad \text{and}\quad \mu_1+\cdots+\mu_i \leq \lambda_1+\cdots+\lambda_i,\quad \forall i=1,2,\dots.
\end{aligned}
\end{equation}
Then $\mu < \lambda$ means $\mu \leq \lambda$ and $\mu \ne \lambda$. Note that the dominance ordering is not a total ordering.

\paragraph{Symmetric polynomials.}
We are interested in symmetric polynomials of $n$-variable $x=(x_1,\dots, x_n)$.
We first introduce elementary and completely symmetric polynomials by
\begin{align}
\sum_{r=0}^\infty z^r e_r(x_1,\dots,x_n)&=\prod_{i=1}^n (1+z x_i),\\
\sum_{r=0}^\infty z^r h_r(x_1,\dots,x_n)&=\prod_{i=1}^n \frac{1}{1-z x_i}.
\end{align}
Obviously, the elementary symmetric polynomials are non-trivial only for $r \leq n$.
Also, power sum symmetric polynomials are defined by
\begin{align}
p_r(x_1,\dots, x_n)=\sum_{i=1}^n x_i^r.
\end{align}
To define the Macdonald polynomials, the following monomial symmetric polynomial is important:
\begin{equation}
\begin{aligned}
m_\lambda(x_1,\dots, x_n)=\sum_{\alpha \in S_n.\lambda} x^\alpha,
\end{aligned}
\end{equation}
where $\alpha=(\alpha_1,\dots, \alpha_n)$ in the sum runs over all distinct permutations of the partition $\lambda=(\lambda_1,\dots, \lambda_n)$, and $x^\alpha=x_1^{\alpha_1}\cdots x_n^{\alpha_n}$.
We show some explicit forms for $n=3$:
\begin{equation}
\begin{aligned}
m_{(3)}(x_1, x_2, x_3)&=x_1^3+x_2^3+x_3^3,\\
m_{(2,1)}(x_1, x_2, x_3)&=x_1^2x_2+x_1 x_2^2+x_1^2 x_3+x_1 x_3^2+x_2^2x_3+x_2 x_3^2,\\
m_{(1^3)}(x_1, x_2, x_3)&=x_1x_2x_3.
\end{aligned}
\end{equation}
Clearly, we have $m_{(r)}(x)=p_r(x)$ and $m_{(1^r)}(x)=e_r(x)$.

\paragraph{Macdonald polynomials.}
Following \cite{Noumi}, we introduce the Macdonald polynomials. Let us consider the following $q$-difference operator:
\begin{equation}
\begin{aligned}
D_x=\sum_{i=1}^n \prod_{1\leq j(\ne i) \leq n} \frac{tx_i-x_j}{x_i-x_j}T_{q, x_i},
\end{aligned}
\end{equation}
where 
\begin{equation}
\begin{aligned}
T_{q, x_i}f(x_1, \dots, x_i, \dots, x_n)=f(x_1,\dots, qx_i,\dots, x_n).
\end{aligned}
\end{equation}
The Macdonald polynomial
\begin{equation}
\begin{aligned}
P_\lambda(x;q,t)=m_\lambda(x)+\sum_{\mu<\lambda} u_\mu^\lambda(q,t) m_\mu(x)
\end{aligned}
\end{equation}
is defined as an eigenfunction of $D_x$. More precisely, it satisfies the following eigenvalue equation,
\begin{equation}
\begin{aligned}
D_x P_\lambda(x;q,t)=d_\lambda(q,t) P_\lambda(x;q,t),
\end{aligned}
\end{equation}
where the eigenvalue is given by
\begin{equation}
\begin{aligned}
d_\lambda(q,t)=\sum_{i=1}^n t^{n-i}q^{\lambda_i}.
\end{aligned}
\end{equation}
For each partition $\lambda$, the Macdonald polynomial $P_\lambda(x;q,t)$ is uniquely fixed by this definition.
For the reader's convenience, we show its explicit forms for $n \geq 3$ and $|\lambda|\leq 3$:
\begin{equation}
\begin{aligned}
P_{(1)}=m_{(1)},\qquad P_{(2)}=m_{(2)}+\frac{(1-q^2)(1-t)}{(1-q)(1-qt)}m_{(1^2)},\qquad P_{(1^2)}=m_{(1^2)},
\end{aligned}
\end{equation}
and
\begin{equation}
\begin{aligned}
P_{(3)}&=m_{(3)}+\frac{(1-q^3)(1-t)}{(1-q)(1-q^2 t)}m_{(2,1)}+\frac{(1-q^2)(1-q^3)(1-t)^2}{(1-q)^2(1-qt)(1-q^2 t)}m_{(1^3)},\\
P_{(2,1)}&=m_{(2,1)}+\frac{(1-t)(2+q+t+2qt)}{1-qt^2}m_{(1^3)}, \\
P_{(1^3)}&=m_{(1^3)}.
\end{aligned}
\end{equation}
By definition, we have
\begin{equation}
\begin{aligned}
P_{(1^r)}(x;q,t)=m_{(1^r)}(x)=e_r(x).
\end{aligned}
\end{equation}
The polynomials $P_{(r)}(x;q,t)$ are generated by
\begin{align}
\prod_{i=1}^n \frac{(tx_i y;q)_\infty}{(x_i y;q)_\infty}&=\sum_{r=0}^\infty g_r(x;q,t) y^r,\label{eq:gen-g}\\
P_{(r)}(x;q,t)&=\frac{(q;q)_r}{(t;q)_r}g_r(x;q,t).
\end{align}
The Macdonald polynomials have the two parameters $(q,t)$, and there are various interesting specializations.
The case $t=q$ is particularly important. In this case, it is well-known that the Macdonald polynomials reduce to the Schur polynomials, 
\begin{equation}
\begin{aligned}
P_\lambda(x;q,q)=s_\lambda(x),
\end{aligned}
\end{equation}
where the Schur polynomials are defined by
\begin{equation}
\begin{aligned}
s_\lambda(x)=\frac{\det( x_i^{\lambda_j+n-j})_{1\leq i,j \leq n}}{\det( x_i^{n-j})_{1\leq i,j \leq n}}.
\end{aligned}
\end{equation}
Also in the limit $t\to 1$, the Macdonald polynomials reduce to the monomial symmetric polynomials: $P_\lambda(x;q,1)=m_\lambda(x)$.
In $q \to 0$, $P_\lambda(x;0,t)$ is the Hall-Littlewood polynomials. When $t \to 0$, $P_\lambda(x;q,0)$ is called the $q$-Whittaker polynomials. See Fig.~1.1 in \cite{Noumi} for other specializations.

The Macdonald polynomials are symmetric orthogonal polynomials. The weight function of them is given by
\begin{equation}
\begin{aligned}
w(x)=\prod_{1\leq i \ne j \leq n}\frac{(x_i/x_j;q)_\infty}{(tx_i/x_j;q)_\infty}.
\end{aligned}
\end{equation}
The orthogonality relation is
\begin{equation}
\begin{aligned}
\frac{1}{n!}\oint_{\mathbb{T}^n} \prod_{i=1}^n \frac{dx_i}{2\pi i x_i} w(x) P_\lambda(x;q,t)P_\mu(x^{-1};q,t)=\delta_{\lambda,\mu}\cN_{\lambda, n},
\end{aligned}
\end{equation}
where $\mathbb{T}^n=\{ (x_1,\dots,x_n) \in \mathbb{C}^n|\; |x_i|=1 \}$ and
\begin{equation}
\begin{aligned}
\cN_{\lambda,n}=\prod_{1 \leq i<j \leq n} \frac{(t^{j-i}q^{\lambda_i-\lambda_j+1};q)_\infty (t^{j-i}q^{\lambda_i-\lambda_j};q)_\infty}{(t^{j-i+1}q^{\lambda_i-\lambda_j};q)_\infty (t^{j-i-1}q^{\lambda_i-\lambda_j+1};q)_\infty}.
\end{aligned}
\end{equation}
The orthogonality is shown by the self-adjointness of the difference operator $D_x$, but a derivation of the formula on the norm $\cN_{\lambda, n}$ is highly non-trivial.

In our analysis, the Cauchy formula plays a crucial role. Let us define
\begin{equation}
\begin{aligned}
\Pi(x,y; q,t)=\prod_{i=1}^n \prod_{j=1}^m \frac{(t x_i y_j;q)_\infty}{(x_i y_j; q)_\infty}.
\end{aligned}
\end{equation}
The Cauchy formula claims that
\begin{equation}
\begin{aligned}
\Pi(x,y; q,t)
=\sum_{\ell(\lambda) \leq \min(n,m)} b_{\lambda} P_\lambda(x;q,t) P_\lambda(y;q,t),
\end{aligned}
\end{equation}
where
\begin{equation}
\begin{aligned}
b_\lambda=\prod_{1 \leq i \leq j \leq \ell(\lambda)} \frac{(t^{j-i+1}q^{\lambda_i-\lambda_j};q)_{\lambda_j-\lambda_{j+1}}}{(t^{j-i}q^{\lambda_i-\lambda_j+1};q)_{\lambda_j-\lambda_{j+1}}}.
\end{aligned}
\end{equation}
When $q=t$, the Cauchy formula reduces to
\begin{equation}
\begin{aligned}
\prod_{i=1}^n \prod_{j=1}^m \frac{1}{1-x_i y_j}
=\sum_{\ell(\lambda) \leq \min(n,m)} s_\lambda(x) s_\lambda(y).
\end{aligned}
\end{equation}
A key point to show the Cauchy formula is that $\Pi(x,y;q,t)$ is a kernel function of the difference operators $D_x$ and $D_y$:
\begin{equation}
\begin{aligned}
D_x \Pi(x,y;q,t)=D_y \Pi(x,y;q,t).
\end{aligned}
\end{equation}

In the analysis of line operator indices, we need the Pieri formula. Let us explain it.
Since the Macdonald polynomials form a basis of symmetric polynomials, the product of two Macdonald polynomials are also expanded by the Macdonald polynomials:
\begin{equation}
\begin{aligned}
P_\lambda(x;q,t)P_\rho(x;q,t)=\sum_{\mu} c_{\lambda\rho}^\mu(q,t) P_\mu(x;q,t).
\end{aligned}
\end{equation}
When $t=q$, $c_{\lambda\rho}^\mu=c_{\lambda\rho}^\mu(q,q)$ is nothing but the Littlewood-Richardson coefficient for the Schur polynomials. Unlike the Schur polynomials, the coefficient $c_{\lambda\rho}^\mu(q,t)$ for the Macdonald polynomials are much more complicated. Fortunately, for $\rho=(1^r)$, the coefficient is explicitly known. This is referred to as the Pieri formula:
\begin{equation}
\begin{aligned}
e_r(x)P_\lambda(x;q,t)=\sum_{\mu \in V_{n}^r(\lambda)} \psi_{\mu/\lambda}'(q,t) P_\mu(x;q,t)
\end{aligned}
\end{equation}
where
\begin{align}
V_{n}^r(\lambda)&=\{ \mu \vdash |\lambda|+r \; | \; \ell(\mu) \leq n \;\; \text{and} \;\; \mu/\lambda\;\; \text{is a vertical strip} \},\label{eq:Vnr}\\
\psi_{\mu/\lambda}(q,t)&=\prod_{1\leq i \leq j \leq \ell(\lambda)} \frac{(t^{j-i+1}q^{\lambda_i-\lambda_j};q)_{\mu_i-\lambda_i}(t^{j-i}q^{\lambda_i-\mu_{j+1}+1};q)_{\mu_i-\lambda_i}}{(t^{j-i}q^{\lambda_i-\lambda_j+1};q)_{\mu_i-\lambda_i}(t^{j-i+1}q^{\lambda_i-\mu_{j+1}};q)_{\mu_i-\lambda_i}},\\
\psi_{\mu/\lambda}'(q,t)&=\psi_{\mu'/\lambda'}(t,q).
\end{align}
There is also another type of the Pieri formula for $\rho=(r)$:
\begin{equation}
\begin{aligned}
g_r(x;q,t)P_\lambda(x;q,t)=\sum_{\mu \in H_{n}^r(\lambda)} \varphi_{\mu/\lambda}(q,t) P_\mu(x;q,t),
\end{aligned}
\label{eq:another-Pieri}
\end{equation}
where
\begin{align}
H_{n}^r(\lambda)&=\{ \mu \vdash |\lambda|+r \; | \; \ell(\mu) \leq n \;\; \text{and} \;\; \mu/\lambda\;\; \text{is a horizontal strip} \},\label{eq:Vnr}\\
\varphi_{\mu/\lambda}(q,t)&=\prod_{1\leq i \leq j \leq \ell(\mu)} \frac{(t^{j-i+1}q^{\mu_i-\mu_j};q)_{\mu_j-\lambda_j}(t^{j-i}q^{\lambda_i-\mu_{j+1}+1};q)_{\mu_{j+1}-\lambda_{j+1}}}{(t^{j-i}q^{\mu_i-\mu_j+1};q)_{\mu_j-\lambda_j}(t^{j-i+1}q^{\lambda_i-\mu_{j+1}};q)_{\mu_{j+1}-\lambda_{j+1}}}.
\end{align}
For example, if $n=4$, $r=2$ and $\lambda=(3,1,1)$, then $V_{4}^2((3,1,1))$ and $H_{4}^2((3,1,1))$ are explicitly given by
\begin{equation}
\begin{aligned}
V_{4}^2((3,1,1))&=\{(4,2,1), (4,1,1,1), (3,2,2), (3,2,1,1) \},\\
H_{4}^2((3,1,1))&=\{(5,1,1), (4,2,1), (4,1,1,1), (3,3,1), (3,2,1,1) \}.
\end{aligned}
\end{equation}

\section{Half-indices of interfaces}\label{sec:interface}
In this appendix, we show additional exact results on half-indices of the $U(N)|U(M)$ interface, introduced in \cite{Gaiotto:2019jvo}.
Without loss of generality, we can assume $N \leq M$. 

The matrix integral for the half-index for NS5-type interface between $U(N)$ and $U(M)$ gauge theories is given by
\begin{equation}
\begin{aligned}
&\mathbb{II}_{\cN}^{U(N)|U(M)}=\frac{1}{N!M!} \frac{(q;q)_\infty^{N+M}}{(t;q)_\infty^{N+M}}
\oint_{\mathbb{T}^N} \prod_{i=1}^N \frac{dx_i}{2\pi i x_i} \prod_{1 \leq i \ne j\leq N} \frac{(x_i/x_j;q)_\infty}{(tx_i/x_j;q)_\infty} \\
&\times \oint_{\mathbb{T}^M} \prod_{i=1}^M \frac{dy_i}{2\pi i y_i} \prod_{1 \leq i \ne j\leq M} \frac{(y_i/y_j;q)_\infty}{(ty_i/y_j;q)_\infty}
\prod_{i=1}^N \prod_{j=1}^M \frac{(tu^{1/2}x_i/y_j;q)_\infty (tu^{1/2}y_j/x_i;q)_\infty}{(u^{1/2}x_i/y_j;q)_\infty (u^{1/2}y_j/x_i;q)_\infty}.
\end{aligned}
\end{equation}
We use the Cauchy formula as
\begin{equation}
\begin{aligned}
\prod_{i=1}^N \prod_{j=1}^M \frac{(tu^{1/2}x_i/y_j;q)_\infty}{(u^{1/2}x_i/y_j;q)_\infty}
&=\sum_{\ell(\lambda) \leq N } u^{|\lambda|/2}b_{\lambda} P_\lambda(x;q,t)P_\lambda(y^{-1};q,t),\\
\prod_{i=1}^N \prod_{j=1}^M \frac{(tu^{1/2}y_ j/x_i;q)_\infty}{(u^{1/2}y_j/x_i;q)_\infty}
&=\sum_{\ell(\mu) \leq N } u^{|\mu|/2}b_{\mu} P_\mu(x^{-1};q,t)P_\mu(y;q,t).
\end{aligned}
\end{equation}
Then we can perform the torus integrals:
\begin{equation}
\begin{aligned}
\mathbb{II}_{\cN}^{U(N)|U(M)}&=\frac{(q;q)_\infty^{N+M}}{(t;q)_\infty^{N+M}}\sum_{\ell(\lambda) \leq N }\sum_{\ell(\mu) \leq N }
u^{|\lambda|/2+|\mu|/2}b_{\lambda}b_{\mu} \delta_{\lambda,\mu}\cN_{\lambda,N}\delta_{\mu,\lambda}\cN_{\mu,M}\\
&=\frac{(q;q)_\infty^{N+M}}{(t;q)_\infty^{N+M}}\sum_{\ell(\lambda) \leq N } u^{|\lambda|}b_{\lambda}\cN_{\lambda,N}b_{\lambda}\cN_{\lambda,M}.
\end{aligned}
\end{equation}
Using \eqref{eq:bN}, we obtain
\begin{equation}
\begin{aligned}
\mathbb{II}_{\cN}^{U(N)|U(M)}=\frac{(q;q)_\infty^2}{(t;t)_N(t^Nq;q)_\infty (t;t)_M (t^Mq;q)_\infty}
\sum_{\ell(\lambda) \leq N } u^{|\lambda|} \prod_{i=1}^{\ell(\lambda)} \frac{(t^{N-i+1};q)_{\lambda_i}(t^{M-i+1};q)_{\lambda_i}}{(t^{N-i}q;q)_{\lambda_i}(t^{M-i}q;q)_{\lambda_i}}.
\end{aligned}
\end{equation}
This result is equivalent to the previous conjecture in \cite{Hatsuda:2024uwt}.

On the other hand, the matrix integral of the half-index of the D5-type $U(N)|U(M)$ interface is given by
\begin{equation}
\begin{aligned}
\mathbb{II}_{\cD'}^{U(N)|U(M)}&=\frac{1}{N!} \frac{(q;q)_\infty^{N}(tu;q)_\infty^N}{(t;q)_\infty^{N}(u;q)_\infty^N}
\oint_{\mathbb{T}^N} \prod_{i=1}^N \frac{dx_i}{2\pi i x_i} \prod_{1 \leq i \ne j\leq N} \frac{(x_i/x_j;q)_\infty(tux_i/x_j;q)_\infty}{(tx_i/x_j;q)_\infty(ux_i/x_j;q)_\infty}\\
&\quad\times \prod_{k=1}^{M-N} \frac{(t^{k-1}q;q)_\infty}{(t^k;q)_\infty} \prod_{i=1}^N \frac{(t^{(M-N+1)/2}ux_i;q)_\infty(t^{(M-N+1)/2}ux_i^{-1};q)_\infty}{(t^{(M-N+1)/2}x_i;q)_\infty(t^{(M-N+1)/2}x_i^{-1};q)_\infty}.
\end{aligned}
\end{equation}
The evaluation of this integral turns out to be more involved. We rewrite it as
\begin{equation}
\begin{aligned}
\mathbb{II}_{\cD'}^{U(N)|U(M)}&=\frac{(q;q)_\infty^{N}}{(u;q)_\infty^N}\prod_{k=1}^{M-N} \frac{(t^{k-1} q;q)_\infty}{(t^k;q)_\infty}\frac{1}{N!}
\oint_{\mathbb{T}^N} \prod_{i=1}^N \frac{dx_i}{2\pi i x_i} \prod_{1 \leq i \ne j\leq N} \frac{(x_i/x_j;q)_\infty}{(ux_i/x_j;q)_\infty}\\
&\quad\times \prod_{i,j=1}^N \frac{(tux_i/x_j;q)_\infty}{(tx_i/x_j;q)_\infty} \prod_{i=1}^N \frac{(t^{(M-N+1)/2}ux_i;q)_\infty(t^{(M-N+1)/2}ux_i^{-1};q)_\infty}{(t^{(M-N+1)/2}x_i;q)_\infty(t^{(M-N+1)/2}x_i^{-1};q)_\infty}.
\end{aligned}
\end{equation}
Note that we need to consider the Macdonald polynomials with two parameters $(q,u)$.
We use the Cauchy formula:
\begin{equation}
\begin{aligned}
\prod_{i,j=1}^N \frac{(tux_i/x_j;q)_\infty}{(tx_i/x_j;q)_\infty}=\sum_{\ell(\lambda)\leq N}t^{|\lambda|}b_\lambda P_\lambda(x;q,u)P_\lambda(x^{-1};q,u),
\end{aligned}
\end{equation}
and the generating function \eqref{eq:gen-g}:
\begin{equation}
\begin{aligned}
\prod_{i=1}^N \frac{(t^{(M-N+1)/2}ux_i;q)_\infty}{(t^{(M-N+1)/2}x_i;q)_\infty}
=\sum_{r=0}^\infty g_r(x;q,u) t^{(M-N+1)r/2}.
\end{aligned}
\end{equation}
Moreover we use another Pieri formula \eqref{eq:another-Pieri}. We finally obtain
\begin{equation}
\begin{aligned}
\mathbb{II}_{\cD'}^{U(N)|U(M)}=\frac{(q;q)_\infty^2}{(t;t)_{M-N}(t^{M-N}q,q)_\infty (u;u)_N (u^N q;q)_\infty}
\sum_{\ell(\lambda) \leq N} \sum_{r=0}^\infty t^{|\lambda|+(M-N+1)r}\\
\times \sum_{\mu \in H_N^r(\lambda)} \varphi_{\mu/\lambda}(q,u)\psi_{\mu/\lambda}(q,u)\prod_{i=1}^{\ell(\mu)}\frac{(u^{N-i+1};q)_{\mu_i}}{(u^{N-i}q;q)_{\mu_i}},
\end{aligned}
\end{equation}
where we have used
\begin{equation}
\begin{aligned}
\prod_{k=1}^{M-N} \frac{(t^{k-1} q;q)_\infty}{(t^k;q)_\infty}=\frac{(q;q)_\infty}{(t;t)_{M-N}(t^{M-N}q;q)_\infty}.
\end{aligned}
\end{equation}
For $q=tu$, these two half-indices are exactly the same,
\begin{equation}
\begin{aligned}
\mathbb{II}_{\cN}^{U(N)|U(M)}=\mathbb{II}_{\cD'}^{U(N)|U(M)}\qquad (q=tu)
\end{aligned}
\end{equation}
To check it, we consider a slice $t=\beta q^{1/2}$ and $u=\beta^{-1}q^{1/2}$ for instance, and expand both indices around $q=0$.
We find a perfect agreement.

\bibliographystyle{amsmod}
\bibliography{Indices}

\ifx\undefined\bysame
\newcommand{\bysame}{\leavevmode\hbox to3em{\hrulefill}\,}
\fi
\begin{thebibliography}{10}

\bibitem{Romelsberger:2005eg}
C.~Romelsberger, {\em {Counting chiral primaries in $\mathcal{N}=1$, $d=4$
  superconformal field theories}}, Nucl. Phys. B {\bf 747} (2006) 329--353,
  {\tt arXiv:hep-th/0510060}.

\bibitem{Kinney:2005ej}
J.~Kinney, J.~M. Maldacena, S.~Minwalla and S.~Raju, {\em {An Index for 4
  dimensional super conformal theories}}, Commun. Math. Phys. {\bf 275} (2007)
  209--254, {\tt arXiv:hep-th/0510251}.

\bibitem{Gadde:2011ik}
A.~Gadde, L.~Rastelli, S.~S. Razamat and W.~Yan, {\em {The 4d Superconformal
  Index from $q$-deformed 2d Yang-Mills}}, Phys. Rev. Lett. {\bf 106} (2011)
  241602, {\tt arXiv:1104.3850} {\tt [hep-th]}.

\bibitem{Gadde:2011uv}
A.~Gadde, L.~Rastelli, S.~S. Razamat and W.~Yan, {\em {Gauge Theories and
  Macdonald Polynomials}}, Commun. Math. Phys. {\bf 319} (2013) 147--193, {\tt
  arXiv:1110.3740} {\tt [hep-th]}.

\bibitem{Spiridonov:2010qv}
V.~P. Spiridonov and G.~S. Vartanov, {\em {Superconformal indices of ${\mathcal
  N}=4$ SYM field theories}}, Lett. Math. Phys. {\bf 100} (2012) 97--118, {\tt
  arXiv:1005.4196} {\tt [hep-th]}.

\bibitem{Dolan:2007rq}
F.~A. Dolan, {\em {Counting BPS operators in $\mathcal{N}=4$ SYM}}, Nucl. Phys.
  B {\bf 790} (2008) 432--464, {\tt arXiv:0704.1038} {\tt [hep-th]}.

\bibitem{Gaiotto:2020vqj}
D.~Gaiotto and J.~Abajian, {\em {Twisted M2 brane holography and sphere
  correlation functions}}, {\tt arXiv:2004.13810} {\tt [hep-th]}.

\bibitem{Gaiotto:2021xce}
D.~Gaiotto and J.~H. Lee, {\em {The giant graviton expansion}}, JHEP {\bf 08}
  (2024) 025, {\tt arXiv:2109.02545} {\tt [hep-th]}.

\bibitem{Hatsuda:2022xdv}
Y.~Hatsuda and T.~Okazaki, {\em {$\mathcal{N}= 2^*$ Schur indices}}, JHEP {\bf
  01} (2023) 029, {\tt arXiv:2208.01426} {\tt [hep-th]}.

\bibitem{Arai:2019xmp}
R.~Arai and Y.~Imamura, {\em {Finite $N$ Corrections to the Superconformal
  Index of S-fold Theories}}, PTEP {\bf 2019} (2019) 083B04, {\tt
  arXiv:1904.09776} {\tt [hep-th]}.

\bibitem{Arai:2020qaj}
R.~Arai, S.~Fujiwara, Y.~Imamura and T.~Mori, {\em {Schur index of the ${\cal
  N}=4$ $U(N)$ supersymmetric Yang-Mills theory via the AdS/CFT
  correspondence}}, Phys. Rev. D {\bf 101} (2020) 086017, {\tt
  arXiv:2001.11667} {\tt [hep-th]}.

\bibitem{Imamura:2021ytr}
Y.~Imamura, {\em {Finite-$N$ superconformal index via the AdS/CFT
  correspondence}}, PTEP {\bf 2021} (2021) 123B05, {\tt arXiv:2108.12090} {\tt
  [hep-th]}.

\bibitem{Fulton}
W.~Fulton and J.~Harris, {\em {Representation theory}}, Springer-Verlag, New
  York, 1991.

\bibitem{Gaiotto:2019jvo}
D.~Gaiotto and T.~Okazaki, {\em {Dualities of Corner Configurations and
  Supersymmetric Indices}}, JHEP {\bf 11} (2019) 056, {\tt arXiv:1902.05175}
  {\tt [hep-th]}.

\bibitem{Macdonald}
I.~G. Macdonald, {\em {Symmetric Functions and Hall Polynomials}}, Clarendon
  Press, 1998.

\bibitem{Noumi}
M.~Noumi, {\em Macdonald polynomials}, Springer Singapore, 2023.

\bibitem{Bourdier:2015wda}
J.~Bourdier, N.~Drukker and J.~Felix, {\em {The exact Schur index of
  $\mathcal{N}=4$ SYM}}, JHEP {\bf 11} (2015) 210, {\tt arXiv:1507.08659} {\tt
  [hep-th]}.

\bibitem{Hatsuda:2024uwt}
Y.~Hatsuda, H.~Lin and T.~Okazaki, {\em {Giant graviton expansions and ETW
  brane}}, JHEP {\bf 09} (2024) 181, {\tt arXiv:2405.14564} {\tt [hep-th]}.

\bibitem{Hatsuda:2023iwi}
Y.~Hatsuda and T.~Okazaki, {\em {Exact $\mathcal{N}=2^*$ Schur line defect
  correlators}}, JHEP {\bf 06} (2023) 169, {\tt arXiv:2303.14887} {\tt
  [hep-th]}.

\bibitem{Hatsuda:2023imp}
Y.~Hatsuda and T.~Okazaki, {\em {Large $N$ and large representations of Schur
  line defect correlators}}, JHEP {\bf 01} (2024) 096, {\tt arXiv:2309.11712}
  {\tt [hep-th]}.

\bibitem{Imamura:2024zvw}
Y.~Imamura, A.~Sei and D.~Yokoyama, {\em {Giant graviton expansion for general
  Wilson line operator indices}}, JHEP {\bf 09} (2024) 202, {\tt
  arXiv:2406.19777} {\tt [hep-th]}.

\bibitem{Murthy:2022ien}
S.~Murthy, {\em {Unitary matrix models, free fermions, and the giant graviton
  expansion}}, Pure Appl. Math. Quart. {\bf 19} (2023) 299--340, {\tt
  arXiv:2202.06897} {\tt [hep-th]}.

\bibitem{Beccaria:2023zjw}
M.~Beccaria and A.~Cabo-Bizet, {\em {On the brane expansion of the Schur
  index}}, JHEP {\bf 08} (2023) 073, {\tt arXiv:2305.17730} {\tt [hep-th]}.

\bibitem{Beccaria:2023hip}
M.~Beccaria and A.~Cabo-Bizet, {\em {Large black hole entropy from the giant
  brane expansion}}, JHEP {\bf 04} (2024) 146, {\tt arXiv:2308.05191} {\tt
  [hep-th]}.

\bibitem{Imamura:2024lkw}
Y.~Imamura, {\em {Giant Graviton Expansions for the Line Operator Index}}, PTEP
  {\bf 2024} (2024) 063B03, {\tt arXiv:2403.11543} {\tt [hep-th]}.

\bibitem{Beccaria:2024oif}
M.~Beccaria, {\em {Schur line defect correlators and giant graviton
  expansion}}, JHEP {\bf 06} (2024) 088, {\tt arXiv:2403.14553} {\tt [hep-th]}.

\bibitem{Imamura:2024pgp}
Y.~Imamura and M.~Inoue, {\em {Brane expansions for anti-symmetric line
  operator index}}, JHEP {\bf 08} (2024) 020, {\tt arXiv:2404.08302} {\tt
  [hep-th]}.

\bibitem{Dolan:2008qi}
F.~A. Dolan and H.~Osborn, {\em {Applications of the Superconformal Index for
  Protected Operators and $q$-Hypergeometric Identities to $\mathcal{N}=1$ Dual
  Theories}}, Nucl. Phys. B {\bf 818} (2009) 137--178, {\tt arXiv:0801.4947}
  {\tt [hep-th]}.

\bibitem{Gang:2012yr}
D.~Gang, E.~Koh and K.~Lee, {\em {Line Operator Index on $S^{1}\times S^{3}$}},
  JHEP {\bf 05} (2012) 007, {\tt arXiv:1201.5539} {\tt [hep-th]}.

\bibitem{Sei:2023fjk}
A.~Sei, {\em {Character Expansion Methods for $\mathrm{USp}(2N)$,
  $\mathrm{SO}(n)$, and $\mathrm{O}(n)$ using the Characters of the Symmetric
  Group}}, {\tt arXiv:2303.03674} {\tt [hep-th]}.

\bibitem{Du:2023kfu}
B.-n. Du, M.-x. Huang and X.~Wang, {\em {Schur indices for $\mathcal{N}=4$
  super-Yang-Mills with more general gauge groups}}, JHEP {\bf 03} (2024) 009,
  {\tt arXiv:2311.08714} {\tt [hep-th]}.

\bibitem{Fujiwara:2023bdc}
S.~Fujiwara, Y.~Imamura, T.~Mori, S.~Murayama and D.~Yokoyama, {\em {Simple-Sum
  Giant Graviton Expansions for Orbifolds and Orientifolds}}, PTEP {\bf 2024}
  (2024) 023B02, {\tt arXiv:2310.03332} {\tt [hep-th]}.

\bibitem{Hatsuda:2024lcc}
Y.~Hatsuda, H.~Lin and T.~Okazaki, {\em {Orbifold ETW brane and half-indices}},
  JHEP {\bf 12} (2024) 227, {\tt arXiv:2409.16841} {\tt [hep-th]}.

\bibitem{Hatsuda:2025jze}
Y.~Hatsuda, H.~Lin and T.~Okazaki, {\em {$\mathcal{N}=4$ line defect
  correlators of type BCD}}, {\tt arXiv:2502.18110} {\tt [hep-th]}.

\bibitem{Macdonald_2003}
I.~G. Macdonald, {\em {Affine Hecke Algebras and Orthogonal Polynomials}},
  Cambridge University Press, 2003.

\bibitem{Koornwinder1991AskeyWilsonPF}
T.~H. Koornwinder, {\em {Askey-Wilson polynomials for root systems of type
  BC}}, 1991.

\end{thebibliography}

\end{document}